\documentclass[journal]{IEEEtran}
\usepackage{times}  
\usepackage{helvet}  
\usepackage{arydshln}
\usepackage{amsfonts}
\usepackage{mathrsfs}
\usepackage{courier}  
\usepackage[hyphens]{url}  
\usepackage{graphicx} 
\usepackage{caption} 
\usepackage{graphicx}
\usepackage{wrapfig}
\usepackage{amsmath,amssymb} 

\usepackage{times}
\usepackage{epsfig}
\usepackage{graphicx}
\usepackage{amsmath}
\usepackage{amssymb}
\usepackage{booktabs}
\usepackage{color}
\usepackage{listings}
\usepackage{xcolor}
\usepackage{lipsum}
\usepackage{listings}
\usepackage{caption}
\usepackage{colortbl}
\usepackage{multirow}
\usepackage{float}
\usepackage[hidelinks,colorlinks=true,linkcolor=blue,citecolor=blue]{hyperref}

\usepackage{graphicx,scalerel}

\newcommand{\tabincell}[2]{\begin{tabular}{@{}#1@{}}#2\end{tabular}}  

\begin{document}
\title{A Domain-specific Perceptual Metric via Contrastive Self-supervised Representation: Applications on Natural and Medical Images}

\author{Hongwei~Bran Li, Chinmay Prabhakar, Suprosanna Shit, Johannes Paetzold, Tamaz Amiranashvili, Jianguo~Zhang, Daniel~Rueckert,~\IEEEmembership{Fellow}, Juan Eugenio Iglesias, Benedikt Wiestler and~Bjoern Menze,

\thanks{H. B. Li and C. Prabhakar contribute equally to this work.}
\thanks{B. Wiestler and B. Menze jointly supervised this work.}

\thanks{H. B. Li, S. Shit, J. Paetzold, D.~Rueckert, and B. Wiestler are with the Department of Computer Science, Technical University of Munich, Germany.}
\thanks{C. Prabhakar and B. Menze are with the Department of Quantitative Biomedicine, University of Zurich, Switzerland.}
\thanks{J.~Zhang is with the Department of Computer Science and Engineering, Southern University of Science and Technology (SUSTech), Shenzhen, China.}
\thanks{J.E. Iglesias is with Athinoula A. Martinos Center for Biomedical Imaging, Massachusetts General Hospital and Harvard Medical School, Boston, USA}}


\maketitle

\begin{abstract}
Quantifying the perceptual similarity of two images is a long-standing problem in low-level computer vision. The natural image domain commonly relies on supervised learning, e.g., a pre-trained \emph{VGG}, to obtain a latent representation. However, due to domain shift, pre-trained models from the natural image domain might not apply to other image domains, such as medical imaging.  
Notably, in medical imaging, evaluating the perceptual similarity is exclusively performed by specialists trained extensively in diverse medical fields. 
Thus, medical imaging remains devoid of \emph{task-specific}, \emph{objective} {perceptual} measures. 
This work answers the question: Is it necessary to rely on supervised learning to obtain an effective representation that could measure perceptual similarity, or is self-supervision sufficient?  
To understand whether recent contrastive self-supervised representation (CSR) may come to the rescue, we start with natural images and systematically evaluate \emph{CSR} as a metric across numerous contemporary architectures and tasks and compare them with existing methods. 
We find that in the natural image domain, \emph{CSR} behaves on par with the supervised one on several perceptual tests as a metric, and in the medical domain \emph{CSR} better quantifies perceptual similarity concerning the experts' ratings. We also demonstrate that \emph{CSR} can significantly improve the image quality in two image synthesis tasks. Finally, our extensive results suggest that perceptuality is an emergent property of \emph{CSR}, which can be adapted to many image domains without requiring annotations.
\end{abstract}


\IEEEpeerreviewmaketitle

\section{Introduction}
The perceptuality of a representation mainly includes two aspects: (1) the ability to quantify perceptual similarity as a metric, and (2) the effectiveness of perceptual representations as a loss function.
Deep supervised representation has merged to be an effective perceptual metric \cite{zhang2018unreasonable} to quantify perceptual similarity compared to traditional pixel-wise metrics such as $\ell$-2 norm and \emph{SSIM} \cite{wang2004image}.  
They are beneficial as representations for a variety of image regression tasks. For example, representations from the \emph{VGG} architecture \cite{simonyan2014very} have been popularly used for neural style transfer \cite{gatys2016image}, super-resolution \cite{johnson2016perceptual}, segmentation \cite{amir2021understanding}, and image synthesis \cite{dosovitskiy2016generating}. 

\begin{figure}[t]
\centering
\includegraphics[width=0.48\textwidth]{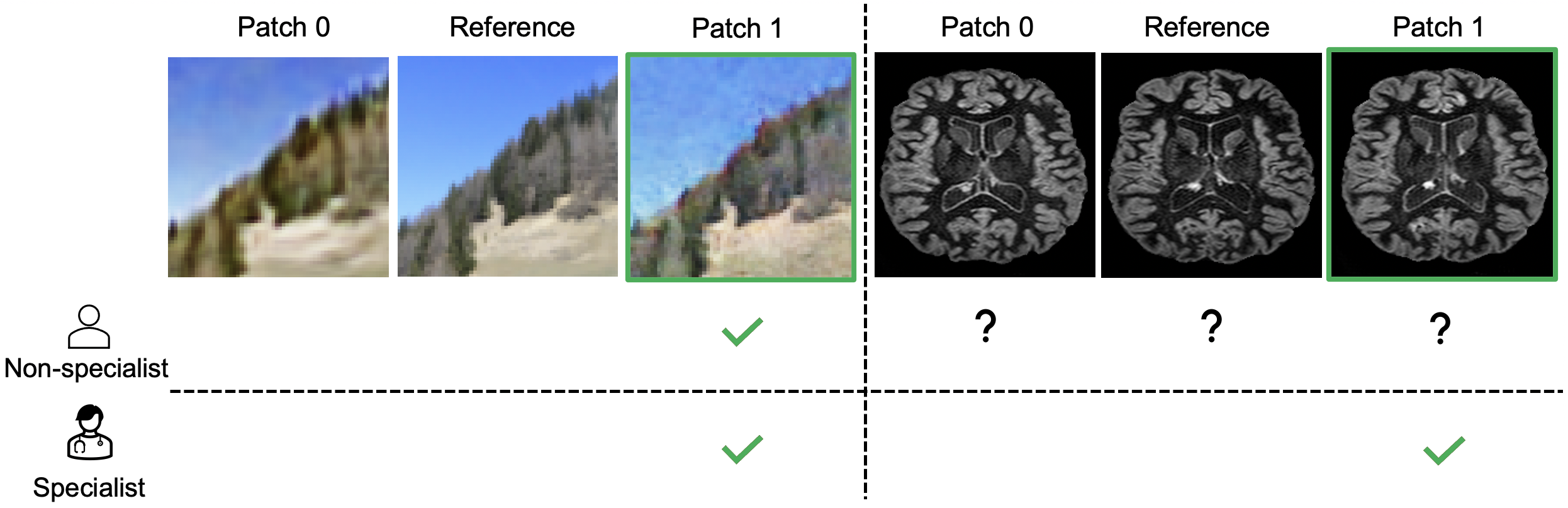}
\caption{\textbf{Quiz: Which patch (left or right) is perceptually closer to the middle patch?} For natural images (\emph{left block}), it is straightforward for most of the humans to compare perceptual similarities. However, medical images (\emph{right block}) are always interpreted by specialists (e.g., radiologists) who would focus on the realism of anatomy and texture. Therefore, it requires \emph{domain-specific} knowledge to compare perceptual similarity in many image domains.}
\label{fig:motivation}
\end{figure}

Although it is relatively straightforward for humans to judge the perceptual similarity between two natural images, some image domains, such as medical imaging and remote sensing \cite{pan2021hyperspectral}, are challenging to interpret without prior knowledge. For example, as shown on the right side in Figure~\ref{fig:motivation}, judging the perceptual similarity of \emph{synthetic} medical images (patch 0 and 1) is exclusively performed by specialists, e.g., neuro-radiologists trained extensively in specific diagnostic tasks. Understanding intricate anatomical structures and subtle pathologies in a given imaging modality, such as magnetic resonance (MR), computed tomography, or ultrasound, is \textbf{domain-knowledge-dependent} and \textbf{task-specific} compared to the natural image domain.

Therefore, an \emph{objective}, \emph{perceptual} metric is required to deal with the diversity of imaging modalities across the medical imaging domain. The problem has been well addressed in the natural image domain by extracting representations from a model pre-trained on \emph{ImageNet} and mapping them to human perceptual rating \cite{deng2009imagenet,zhang2018unreasonable}.
Analogously, for medical images, one naive solution would be training new models in the medical domain in a \emph{supervised} fashion. However, it is resource-consuming and often infeasible to gather large-scale annotated datasets relying on narrative and expensive domain knowledge in the diverse image domains mentioned above. Another solution would be borrowing the available pre-trained models (e.g.~\emph{VGG} \cite{simonyan2014very}, \emph{ResNet} \cite{he2016deep}) to quantify new images from different domains. Nevertheless, it has not been studied how effective the representations learned from \emph{ImageNet} can be used as a perceptual metric for the medical domain due to the large domain shift between them.

With these limitations, we ask whether it is necessary to rely on supervised learning to obtain a good perceptuality of the representation. Or is self-supervision sufficient? We believe that the supervision from labels might not be required because (1) quantifying the perceptual similarity is more structural and implicit than classification tasks. (2) in the natural image domain, contrastive representation learning gradually closes the performance gap between supervised and unsupervised methods. Thus, good representations can be obtained without supervision. Thus, \emph{CSR} is a promising alternative at our disposal for learning more generalizable and effective perceptual representations for the medical domain where labels are difficult to obtain. 
In this paper, we aim to investigate the effectiveness of \emph{CSR} as a perceptual metric and in an auxiliary loss.

\noindent
\textbf{Contributions.} (1) We show that contrastive self-supervised representation behaves on par with supervised ones in quantifying perceptual similarity in the natural image domain. 
The ability of \emph{CSR} to capture perceptual similarity is highly correlated to its representational power, i.e., top-1 accuracy on \emph{ImageNet}. (2) In a visual rating experiment on medical images, we demonstrate that the domain-specific self-supervision-based metric agrees better with the specialists than the one derived from \emph{ImageNet}.  
(3) We introduce a domain-specific perceptual loss to enhance the existing image synthesis framework. We demonstrate that the new perceptual loss consistently improves image quality and the downstream task in two medical image synthesis applications.

\section{Related work}
\subsubsection{Perceptual metrics.} 
Comparing two images by measuring the similarity between their statistics and features is a long-standing problem in computer vision. Classic pixel-wise $\ell$-1 and $\ell$-2 distances, commonly used in regression tasks, are insufficient for assessing structured output due to their assumption of pixel independence. Alternatively, Zhang~\emph{et~al.}~\cite{zhang2018unreasonable} demonstrated the effectiveness of supervised features in correlating human judgment in perceptual tests. The perceptual judgments were based on a two-alternative-forced-choice (2AFC) task, where each test was composed of a reference image and two distortions of the reference (as shown in Fig. \ref{fig:motivation}). Participants were asked to choose which of the two distortions was more similar perceptually to the reference. Following the same experiment design, we will evaluate \emph{CSR} on contemporary architectures as a perceptual metric and compare them against classical and supervised representations-based ones.
\subsubsection{Perceptual representations as loss functions.} 
Perceptual representations consider the discrepancy between the image features instead of the pixel intensity. Using them, a loss term can be formulated by computing the similarity between the representations in style transfer or super-resolution tasks~\cite{gatys2016image,johnson2016perceptual}. 
Instead of using learned representations, Liu \emph{et~al.}~\cite{liu2021generic} argues that a properly randomized VGG can extract generic perceptual features without learning. Although they demonstrated that the random networks could capture the dependencies between multiple levels of variable statistics and enhance baseline methods (e.g., pixel-wise $\ell$-1 norm loss), it is still unclear how effectively they can capture perceptuality compared to supervised ones. 
Similarly, Amir \emph{et al.} \cite{amir2021understanding} proposed an interpretable perceptual loss by computing the maximum mean discrepancy distance between two images with random filters. However, both of them are not optimized for specific domains. Hence it limits their abilities to capture underlying perceptual similarity in challenging scenarios such as depth estimation ~\cite{wang2020cliffnet}. 
Unlike existing work, we propose a \emph{CSR-based} representation that is data-driven, domain-specific, and applicable to any image domain.
\subsubsection{Contrastive self-supervised learning}
The main motivation for introducing \emph{CSR} as a perceptual metric is that it learns useful representations directly from data and is rapidly closing the gap with supervised methods, not only on large computer vision benchmarks \cite{caron2020unsupervised,he2020momentum,chen2020simple,caron2021emerging,zbontar2021barlow} but also in medical imaging \cite{azizi2021big,chaitanya2020contrastive,taleb20203d,zeng2021positional}. The self-supervision is commonly performed by comparing raw or encoded representations from two augmented views. Recently, contrastive learning was employed to assess image quality~\cite{madhusudana2022image}. However, how well the \emph{CSR} correlates with perceptual similarity and how differently they are in capturing perceptual similarity is yet to be explored in both natural and medical image domains.     

\subsubsection{Medical image visual rating and translation.} Interpreting the image quality of medical images is almost exclusively done by specialists. Recent works \cite{li2019diamondgan,kofler2021we} proposed visual ratings in which the radiologists are given random real and synthetic images. Image-to-image synthesis \cite{isola2017image,zhu2017unpaired} approaches in medical imaging can augment available datasets \cite{qasim2020red} and generate missing data \cite{yurt2021mustgan,thomas2021improving}. Mason~\emph{et~al.}~\cite{mason2019comparison} compared objective image quality metrics for synthetic images. This work focuses on quantifying perceptual similarity instead of image quality. To assess the efficacy of the proposed perceptual loss, we perform experiments on two MR datasets comprising image synthesis tasks and follow-up segmentation as a downstream task.

\begin{figure*}[!t]
\centering
\includegraphics[width=1\textwidth]{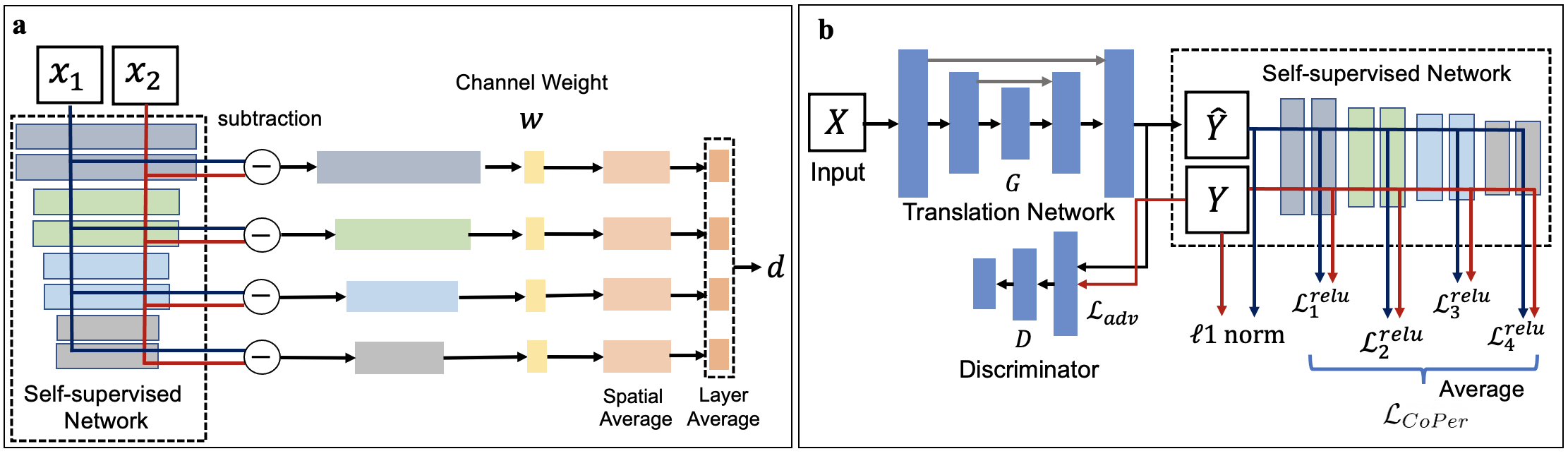}
\caption{\textbf{(a) Perceptual similarity computation:} Given two images $x_{1}$, $x_{2}$ and a pre-trained network, we first compute their multi-scale deep feature maps and normalize the activations in the channel dimension. Next, we compute the difference and scale each channel with a weight $w$. Then, we average across all spatial dimensions and layers to obtain the final distance \emph{d}. \textbf{(b) Image synthesis with a CSR-based perceptual loss:} In an image synthesis framework, we train a generator $G$ and a discriminator $D$ to convert the multi-modal input $X$ into the target modalities $Y$, i.e., $G(X)$$\rightarrow$$Y$. A self-supervised network (e.g., \emph{ResNet} or \emph{VGG}) is used to extract multi-level perceptual features given the real (\emph{Y}) and fake (\emph{$\hat{Y}$}) images and compute the perceptual loss ($\mathcal{L}_{CoPer}$) by averaging the features. Notably, the input \emph{X} and output $Y$ can be with an arbitrary number of channels while the self-supervised network extracts features from each channel of $Y$.}
\label{fig:main_framework}
\end{figure*}

\section{Methods}
In this section, we start from contrastive self-supervised training of domain-specific models. Then, we introduce a self-supervised representation-based perceptual metric. Finally, we make use of the self-supervised representation as a perceptual loss in image synthesis tasks.

\subsection{Contrastive self-supervised pre-training}
The first step is to pre-train a network $f_{\theta}$ on a dataset from a specific domain without supervision. The contrastive
self-supervision is commonly performed by comparing the features from two augmented views. For pre-training on natural images (i.e., \emph{ImageNet-1K}), we directly borrowed the pre-trained networks from existing works \cite{caron2020unsupervised,he2020momentum,chen2020simple,caron2021emerging,zbontar2021barlow}. 

For pre-training on medical images, we adopted one recent approach (\emph{SwAV}) \cite{caron2020unsupervised} and trained the backbones (\emph{ResNet} and \emph{VGG}) on two MRI datasets from scratch. The basic idea of \emph{SwAV} is to compare the encoded feature representations from two augmented views after feature clustering. The reason to choose \emph{SwAV} is that we find a high correlation between its ability to capture perceptual similarity and its representational power, which is to be presented in the results section.

\subsection{\textbf{\emph{CoPer metric}}: a CSR-based perceptual metric}
We set up the contrastive self-supervision-based perceptual metric (\emph{CoPer metric}) by computing the distance of features extracted from multiple levels. Fig.~ \ref{fig:main_framework}(a) shows the steps of computing the perceptual similarity given two images and a pre-trained network following \cite{zhang2018unreasonable}. 
To compute a perceptual similarity $d$ for a pair of images \{x$_{1}$, x$_{2}$\}, given a pre-trained network $\mathscr{F}$, we first extract features from $L$ layers and unit-normalize in the channel dimension. We designate it as $\hat{y}^{l}_{0}$,~$\hat{y}^{l}_{1}~\in~\mathbb{R}^{H_l\times W_l \times C_l}$  for layer $l$. Next, we scale the activations channel-wise by vector $w_{l}$~$\in$~$\mathbb{R}^{C_{l}}$ and compute the $\ell$-2 distance. Finally, we calculate the spatial average and sum channel-wise. An extension of the method to Vision Transformer (ViT) is presented in the Supplementary. 


\label{CoPer}
\subsection{\textbf{\emph{CoPer loss}}: a CSR-based perceptual loss}
In the following, we demonstrate that the perceptual information could be used for medical image synthesis in a paired setting while it also applies for unpaired synthesis without further modification.
\paragraph{Multi-modal image-to-image synthesis.} Given an input set of $m$ image modalities (e.g., multi-contrast MRI): $X = \{x_i|i = 1, ..., m\}$ and a target set of $n$ modalities $Y$. The objective is to learn a translation network $G$ that maps multiple input modalities to multiple target modalities. We assume that all the modalities, i.e., $X$ and $Y$, are spatially aligned. 
We train $G$ to translate $X$ into the target modalities $Y$, i.e., $G(X)$ $\rightarrow$ $Y$. A basic reconstruction loss considering pixel-wise intensity is formulated as: 
\begin{equation} \label{equation_1}
	\mathcal{L}_{\ell1} = \mathbb{E}_{X,Y}[||Y - (G(X))||_{1}]
\end{equation}
	
To enforce the generated images indistinguishable from real ones, we adopt an adversarial loss for discriminator $D$:
\begin{equation} \label{equation_2}
	\mathcal{L}_{adv} = \mathbb{E}_{Y}[log~D(Y)] + \mathbb{E}_{X}[log~(1 - D(G(X)))]
\end{equation}	
	
\paragraph{Image synthesis with the new perceptual loss.}
To incorporate the self-supervised representation as a loss function in image regression tasks, we focus on two image synthesis tasks and propose the framework in Fig.~\ref{fig:main_framework} (b). It includes two components: (i) a GAN-based paired image-to-image synthesis network \cite{isola2017image}, and (ii) a self-supervision-based perceptual loss in addition to a pixel-wise $\ell$1-norm loss. 
We linearly combine the $\ell$1-norm reconstruction loss and the \emph{CoPer} loss during the training. 

We incorporate the pre-trained network as a component of the image synthesis framework to leverage contrastive self-supervised representation as a perceptual loss (nicknamed \emph{CoPer loss}). As shown in Fig.~\ref{fig:main_framework}b), we encourage the output $\hat Y$ and the ground truth $Y$ to have similar perceptual representations as computed by the self-supervised loss network $f_\theta$. Let $\phi_{ij}(X)$ be the $ReLU$ activations of the $i^{th}$ convolutional layer of the network when processing the input $X$ given the $j^{th}$ channel of $Y$. $\phi_{ij}(X)$ is a feature map with a shape of $C_{i} \times H_{i}\times W_{i}$. The feature reconstruction loss is the Euclidean distance between feature representations.

\begin{equation} \label{equation_3}
	\mathcal{L}^{relu}_{ij}(y_{j}, \hat y_{j}) = \frac{1}{C_{i}H_{i}W_{i}} ||\phi_{ij}(y_{j})-\phi_{ij}(\hat y_{j})||_{2}^{2}
	\end{equation}

For each channel of output $Y$, we extract $n$ levels of representations, resulting in the proposed \emph{CoPer} loss: 

\begin{equation} \label{equation_4}
	\mathcal{L}_{CoPer} = \frac{1}{mn}\sum_{j=1}^{m}\sum_{i=1}^{n}	\mathcal{L}^{relu}_{ij}(y_{j}, \hat y_{j}) 
	\end{equation}

\noindent
\emph{Full objective:} The objective functions to optimize $G$ and $D$ respectively are:
\begin{equation} \label{equation_5}
	\mathcal{L}_{G} = \lambda_{1}\mathcal{L}_{\ell1}+(1-\lambda_{1})\mathcal{L}_{CoPer}+\lambda_{2}\mathcal{L}_{adv}
	\end{equation}
\begin{equation} \label{equation_6}
	\mathcal{L}_{D} = -\mathcal{L}_{adv}
	\end{equation}

where $\lambda_{1}$ and $\lambda_{2}$ are hyper-parameters that balance the two reconstruction losses and the adversarial loss.
The effect of $\lambda_{1}$ to balance pixel-wise loss and perceptual loss will be analyzed in the experiments. We use the default setting $\lambda_{2}$ = 10 \cite{isola2017image} in all of our experiments. 

\section{Experiments}
In this section, we first detail the perceptual test setup for both natural and medical images, accompanied by a brief account of the medical data used in our experiments. Next, we describe our self-supervised model training and thereby the image synthesis task using the new perceptual loss. Finally, we go through our findings and look for well-argued answers to our questions posed in the introduction.

\subsection{Datasets and perceptual tests} \label{distance_normalization}
\subsubsection{Natural image dataset and its perceptual test.} The Berkeley-Adobe Perceptual Patch Similarity
(BAPPS) Dataset \cite{zhang2018unreasonable}  is used to evaluate the ability of the methods to capture perceptual similarity. As mentioned in Related Work, the perceptual judgments were based on a two-alternative-forced-choice (2AFC) test. In the \emph{2AFC} test, two distorted images ($x_{1}$, $x_{2}$) and a reference image $x_{R}$ are shown to observers who are asked to choose the distorted image that is closer to the reference. Perceptual metrics are evaluated by measuring the agreement between the distances generated by a network and judgments from observers. The accuracies are reported for five categories of image distortion, including traditional distortion (e.g., random noise and blurring), CNN-based distortion, super-resolution, deblurring, and frame interpolation. 

\subsubsection{Medical datasets for self-supervised pre-training.} We use two brain MRI datasets, including two pathologies: brain tumor and multiple sclerosis (MS).  

\noindent
\textbf{\emph{Brain tumor.}} We use the publicly available BraTS dataset \cite{bakas2018identifying,menze2014multimodal} consisting of four imaging modalities including FLAIR, T1, T2, and T1-c. We use 2D slices of 585 MRI scans, resulting in more than \emph{88k} images after excluding slices with limited content (i.e., the number of foreground pixels are less than 20\% of the image after thresholding). We crop all 2D images to a size of 224~$\times$~224 pixels and rescale their intensities to~[0,~255]. 

\noindent
\textbf{\emph{MS lesions.}} To highlight the potential clinical impact of the proposed method, we also perform self-supervised pre-training on an in-house dataset with MS patients. The set consists of 2D slices from 250 MRI scans acquired with a multi-parametric protocol, including co-registered FLAIR, T1, T2, double inversion recovery (DIR) and contrast-enhanced T1 (T1-c). The images were pre-processed similar to the \emph{BraTS}, resulting in \emph{51k} 2D images. 

\begin{table}[t]
\footnotesize
\setlength{\tabcolsep}{4.4pt}
\def\arraystretch{1}
\centering
\caption{\textbf{Two multi-modal medical datasets for self-supervised pre-training}. For each dataset, 2D slices from two modalities were mixed during the pre-training stage, resulting in  \emph{88k} and \emph{51k} images, respectively.}
\begin{tabular}{l  c  c  c  c  c }
\specialrule{.1em}{0em}{.1em}
{\textbf{Datasets}} & \tabincell{c}{ MRI Modalities} & \tabincell{c}{Image size} &  \tabincell{c}{$\#$ Images} \\ 
\specialrule{.1em}{0em}{.1em}
Brain tumor & FLAIR, T1, T2, T1-c & 192$\times$240$\times$240 & \emph{88k}  \\ 
{MS lesion} & FLAIR, T1, T2, DIR & 192$\times$240$\times$240 & \emph{51k} \\ 
\specialrule{.1em}{0em}{.6em}
\end{tabular}

\end{table}

\begin{table*}[t!]
\centering
\setlength{\tabcolsep}{2.9pt}
\def\arraystretch{0.98}
\caption{\textbf{The effectiveness of self-supervised representation as a perceptual metric on natural images in perceptual tests}. We compared the proposed metric with low-level metrics (e.g., \emph{SSIM}) and supervised representation-based metrics. $\dagger$~denotes \emph{ResNet-50} architecture~\cite{he2016deep}; $\mathsection$ DINO is a \emph{vision transformer}. The three top-performing methods in perceptual similarity judgments for each image distortion are highlighted in blue.}
\begin{tabular}{llcccccc}
\specialrule{.1em}{0em}{.1em}
&{\textbf{Methods}} & \tabincell{c}{traditional \\ distortion} & \tabincell{c}{CNN \\distortion} &\tabincell{c}{super\\resolution}& {deblurring} & \tabincell{c}{frame \\ interpolation} &\textbf{Overall} \\
\specialrule{.1em}{0em}{.1em}
&{$\ell$-1 norm} & {59.94} & {77.77} & {64.67} & {58.20} & {55.02} & 63.12  \\ 
& {SSIM \cite{wang2004image}} & {62.73} & {77.60} & {63.13} & {54.23} & {57.11} & 62.96   \\

\specialrule{.1em}{0em}{.1em}
\parbox[t]{1mm}{\multirow{3}{*}{\rotatebox[origin=c]{90}{\scriptsize random}}} &
{VGG-16 } & 59.86 & 80.84 & 59.06 & 57.73  & 56.97 & 62.89\\ 
& {ResNet-34} &  59.81 & 80.83 & 66.27 & 59.02 & 57.53 & 64.69\\ 
&{ResNet-50\textsuperscript{$\dagger$}} & 59.60 & 80.61 & 66.15 & 58.90 & 57.55 &64.56\\ 

\specialrule{.1em}{.15em}{.1em}
\parbox[t]{1mm}{\multirow{2}{*}{\rotatebox[origin=c]{90}{\scriptsize pretext}}} &
{Colorization\textsuperscript{$\dagger$}} & 59.70 & {80.57}& {{66.79}} & 58.96 & {57.94} &64.79 \\  
& {Jigsaw$\dagger$} & {59.61} & {80.40} & {66.72} & {58.89} & {58.08} &64.74  \\ 

\specialrule{.1em}{0.3em}{.2em}
\parbox[t]{1mm}{\multirow{3}{*}{\rotatebox[origin=c]{90}{\scriptsize supervised}}} &
{VGG-16 } & {68.27} & {80.19} & {67.71} & {57.42} & {61.86} & 67.09  \\ 

& {ResNet-34} & \cellcolor{blue!10}\textbf{71.75} & {80.19} & \cellcolor{blue!10}{{68.87}} & 58.51 & {61.81} & \cellcolor{blue!10}{68.23}  \\ 

&{ResNet-50\textsuperscript{$\dagger$}} & {\cellcolor{blue!10}{71.50}} & {80.47} & \cellcolor{blue!10}{\textbf{69.58}} & {58.15} & \cellcolor{blue!10}{62.08}  & \cellcolor{blue!10}\textbf{68.36} \\ 

\specialrule{.1em}{0.30em}{.2em}
\parbox[t]{1mm}{\multirow{5}{*}{\rotatebox[origin=c]{90}{\scriptsize CSR}}}
& {SimCLR\textsuperscript{$\dagger$} \cite{chen2020simple}} & {60.48} & \cellcolor{blue!10}{81.13} & {66.98} & {{59.03}} & {57.89} & 65.10  \\ 

& {MoCo-v2\textsuperscript{$\dagger$} \cite{he2020momentum}} & {60.07} & {\cellcolor{blue!10}80.95} & {66.91} & \cellcolor{blue!10}{{59.11}} & {58.15} &65.04 \\ 

& {SimSiam\textsuperscript{$\dagger$} \cite{chen2021exploring}} & {61.13} & {80.74} & {66.98} & \cellcolor{blue!10}{59.42} & {58.11} &65.28 \\ 

&{DINO\textsuperscript{$\mathsection$} \cite{caron2021emerging}}& {66.28} & \cellcolor{blue!10}\textbf{82.44} & {66.42} & \cellcolor{blue!10}\textbf{59.61} & \multicolumn{1}{c}{59.51}  & 66.85\\

&\tabincell{l}{Barlow Twins\textsuperscript{$\dagger$} \cite{zbontar2021barlow}}  & {\cellcolor{blue!10}70.94} & {79.63} & \cellcolor{blue!10}{69.15} & {58.16} & \cellcolor{blue!10}{62.00} &  \cellcolor{blue!10}67.98  \\ 

&{SwAV\textsuperscript{$\dagger$} \cite{caron2020unsupervised}}& {69.14} & {79.57} & {68.70} & {57.37}  &  {\cellcolor{blue!10}\textbf{62.16}} &  67.39\\ 

\specialrule{.1em}{0em}{.6em}
\end{tabular}
\label{effectiveness}
\end{table*}

\noindent
\subsubsection{Perceptual rating by two neuroradiologists.} The objective of the rating experiment on medical images is to find out if the perceptual similarity estimated by a network reflects the expert's evaluation. As opposed to the \emph{2AFC} test, which only contains binary ratings, we aim to get finer estimates of the image quality from experts. We design a multi-grade quality evaluation experiment. Two neuroradiologists with 5+ years of professional experience participated in two rating tasks. In each trial, they were provided with two images. One is a reference (real) image from one source MR image domain; the other is a synthetic or real image from the target MR image domain.

We generate two sets of synthetic images named \emph{Set 1} and \emph{Set 2} using \emph{CycleGAN} \cite{zhu2017unpaired} by using different training sets. \emph{Set 1} is generated by the mapping: T1 $\rightarrow$ DIR while \emph{Set 2} is generated by combining T1, T2, and FLAIR to synthesize DIR, i.e., T1+T2+FLAIR~$\rightarrow$~DIR. Notably, the images in \emph{Set 2} are more realistic than \emph{Set 1}. 

Each expert evaluated one set of original images (35 in total) and its corresponding two sets of synthetic images (70 in total) generated by two different inputs.
The paired images were randomly chosen from the generated and real images pool. This particular setup enables the experts to identify perceptual tiny inconsistencies or implausibilities. The experts were asked to rate the quality of the target image via a~{6-star~rating}, where \emph{6 stars} denotes a perfectly plausible image and \emph{1 star} a completely bad quality image.
The ratings are then used to analyze the perceptual similarity generated by different methods. 

\noindent
\subsubsection{Normalization of the perceptual similarity and~the~ratings~from experts.}
Since the perceptual similarity produced by the networks and the experts' ratings are at different ranges, it is essential to normalize them to a certain interval. Let $\mathcal{X}_{R}$ be a set of $n$ real images and $\mathcal{X}_{S}$ be a set of $n$ synthetic images.   
Let $\mathcal{D}$ denote the set of the perceptual similarity computed by comparing $\mathcal{X}_{R}$ and $\mathcal{X}_{S}$: 
$\mathcal{D}=\{d_{i}\}_{i=1}^n$. We normalize $\mathcal{D}$ by dividing by the maximum distance of the set:  
$\mathcal{D}' =\{d'_{i} |d'_{i} = d_{i}/max(\{d_{j}\}_{j=1}^n)\}_{i=1}^n$. 
Similarly, for the image rating from the $k^{th}$ expert, let $\mathcal{R}_{R}^{k}$ and $\mathcal{R}_{S}^{k}$ denote two sets of rating results for the real and synthetic images respectively. We first calculate the \emph{absolute} distance between the two rating sets, denoted as $\mathcal{R}_{abs}^{k}=\{r_{i}^{abs}|r_{i}^{abs}=|r_{i}^R-r_{i}^S|\}_{i=1}^n$. Then we normalize $\mathcal{R}_{abs}^{k}$ by dividing by the maximum distance of the set.

\subsection{Details of model training}
\paragraph{Self-supervised pre-training on medical images.}
We perform contrastive self-supervised pre-training of two modern architectures \emph{ResNet-50} and \emph{VGG} on 2D MRI images. For different translation tasks, the pre-trainings were performed on specific modalities to extract the perceptual representations of target modalities. We consider different imaging modalities (FLAIR \emph{vs.} T2) to be different image domains considering the data heterogeneity within medical domain. The effect of such 'intra'-domain shift will be discussed in the Appendix. Thus, we train individual models for two applications. We adopt the contrastive training strategy from the \emph{SwAV} method \cite{caron2020unsupervised} and train the backbone on medical data from scratch. The training time for 500 epochs of \emph{51k} images is around 48 hours on an Nvidia RTX-3090. We include the training details in the Appendix.

\paragraph{Medical image synthesis tasks and evaluation metrics.}
We performed two image synthesis tasks for the studied two brain diseases. For MS patients, we aim to generate FLAIR images by combining T1 and T2 as the input, while for brain tumor patients, our target is to generate T2 images by using FLAIR and T1 as the input. The two tasks are designed to improve image quality (when FLAIR is with motion) and generate missing T2 modality. We train the models with fixed hyperparameters for all the tasks and fair comparisons. The details of the training are given in the Appendix. We evaluate the quality of synthetic images using: a) direct quantitative metrics, i.e., PSNR \cite{hore2010image}, SSIM, and the newly proposed \emph{CoPer} metric; and b) indirect performance measure from the downstream segmentation task using the synthetic images.

\section{Results}

\begin{figure*}[t]
\centering
\includegraphics[width=0.92\textwidth]{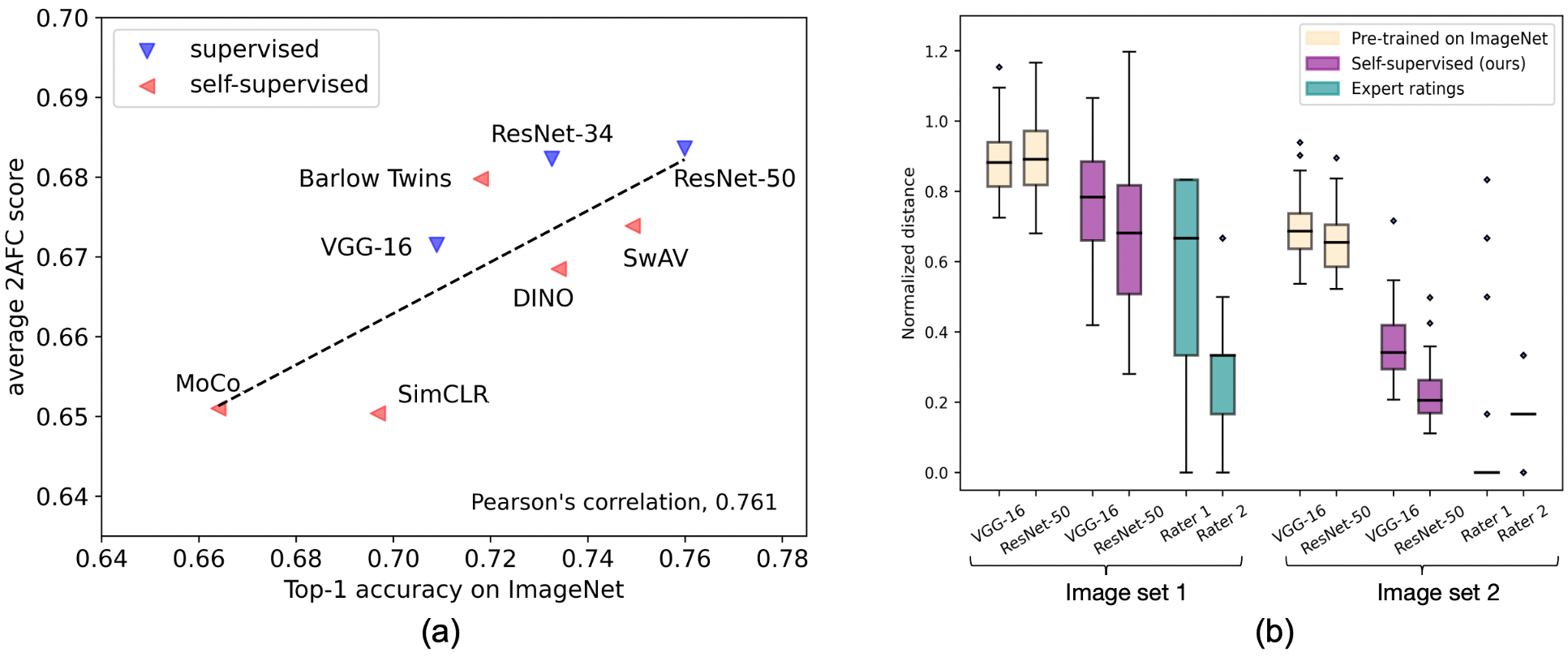}
\vspace{-1mm}
\caption{\textbf{(a) A strong correlation between the representation effectiveness and the ability to capture perceptual similarity.} We correlate the methods' performances on the human perceptual tests with their classification accuracies on \emph{ImageNet}. The \emph{2AFC} scores are averaged across five tasks in Table~\ref{effectiveness}. \textbf{(b) The perceptual similarity computed by \emph{CSR}-based metrics are closer to two experts' ratings than \emph{ImageNet}-derived ones.} We asked two neuroradiologists to rate two sets of synthetic images and their corresponding real images. Then we compute the perceptual similarity by two different methods and architectures. The ratings and distances are normalized to [0, 1]. Set 1 and Set 2 are generated by different inputs, and Set 2 is significantly more realistic than Set 1. We show that the perceptual similarity measured by the \emph{CSR}-based metrics are closer to experts' ratings than the supervised ones pre-trained on \emph{ImageNet}. Especially, they produce \emph{small} perceptual similarity for indistinguishable synthetic images in image set 2.}
\vspace{-1.0mm}
\label{fig:scatter_plot}
\end{figure*}


\vspace{-1mm}
\subsection{How well do self-supervised methods perform perceptual tests on natural images?}
We compare five categories of perceptual metrics on the validation set of \emph{BAPPS}: (1) classical methods including $\ell$-1 norm and structural similarity \cite{wang2004image}, (2) deep networks with random weights, (3) self-trained networks with pre-text tasks, (4) supervised networks trained on \emph{ImageNet}, and (5) contrastive self-supervised methods including recent architectures (\emph{ResNet} and vision transformer). We directly compare the original representations of all methods without further tuning the representations. 

\noindent
\textbf{\emph{Deep features vs. traditional metrics.}}
As shown in Tab.~\ref{effectiveness} (row 1 to 5), we find that \emph{ResNets} and \emph{VGG} with random weights outperform the classical metrics ($\ell1$-norm and \emph{SSIM}); this finding is in line with previous work \cite{zhang2018unreasonable,amir2021understanding}. 

\noindent
\textbf{\emph{Random networks vs. supervised ones.}} Since the random networks are not optimized to specific image domains, they cannot sufficiently capture perceptual similarity compared to supervised (59.60 \emph{vs.} 71.50 with \emph{ResNet-50} in column 2). However, generally they are superior to traditional metrics.

\noindent
\textbf{\emph{Contrastive learning vs. pre-text learning.}} When we directly compare the two categories of self-supervised methods, we found that contrastive learning methods outperform pre-text ones, including colorization~\cite{zhang2016colorful} and solving Jigsaw puzzles~\cite{noroozi2016unsupervised} for most of the image distortion methods. 

\noindent
\textbf{\emph{Contrastive vs. supervised.}}
More importantly, we observe that the self-supervised methods yield comparable performances to the supervised ones in all categories. Especially, in the perceptual test of CNN distortion, \emph{vision transformer} outperforms supervised \emph{ResNet-50} (82.44 \emph{vs.} 80.47), while in frame interpolation, the state-of-the-art self-supervised method \emph{SwAV} (\emph{ResNet-50}) slightly outperform supervised \emph{ResNet-50} (62.16 \emph{vs.} 62.08). These results motivate us to believe that CSR-based methods can be adequate alternatives to supervised ones to capture perceptual similarity. 

Another finding is that \emph{SwAV} and \emph{Barlow Twins} outperform other methods in traditional distortion by a large margin. Traditional distortion includes random noise, spatial shifts, etc., almost the data augmentation schema in the contrastive framework.
We argue that the reasons are related to training strategies -
\emph{SwAV} enforces consistency between the cluster assignments from different augmentations;
\emph{Barlow Twins} learns a compact embedding by measuring the cross-correlation matrix between two embeddings.
However, other methods learn features relatively invariant to these augmentations because they directly minimize the distance between two embeddings.


\begin{table*}[!t]
\centering
\setlength{\tabcolsep}{3pt}
\caption{\textbf{Quantitative comparison of the image quality generated by different loss functions.} For image quality assessment, the scores for the two datasets are averaged across the 2D slices of the validation set. We find that self-supervision-based perceptual loss improves the image quality of synthetic images and downstream segmentation tasks (Dice scores). 
In the segmentation task, the scores for the two datasets are averaged across the validation set consisting of 30 and 50 3D scans, respectively. Dice denotes Dice coefficient which measures the overlap between prediction and reference segmentation annotated by experts. In the  brain tumor segmentation task, \emph{NC}, \emph{D}, and \emph{T} denote three tumor structures \cite{menze2014multimodal}. `Upper bound' denotes the performance of using \emph{real images}, as opposed to synthetic ones. `rand.' denotes `randomized.' }
\begin{tabular}{llllll|llllll}
\specialrule{.1em}{0em}{.1em}
&\multicolumn{1}{c}{\multirow{2}{*}{\textbf{Methods}}} & \multicolumn{4}{c}{MS lesion}  & \multicolumn{6}{c}{Brain tumor} \\ \cline{3-12} & & {PSNR} & {SSIM} & CoPer $\downarrow$ & Dice & \multicolumn{1}{l}{PSNR} & {SSIM} & CoPer $\downarrow$ & Dice$_{NC}$ &Dice$_{D}$ & Dice$_{T}$\\ \specialrule{.1em}{0em}{.1em}
&{$\ell$-1 norm loss} & 32.773 & 0.862  & 2.438 &0.517 & {34.152}  & 0.887   &  2.954 & 0.824 & 0.908 & 0.920 \\
 \midrule
& + rand. ResNet \cite{liu2021generic}& 32.821 & 0.871 & 2.347  & 0.539 & 34.221 &0.889 & 2.877 &0.830  & 0.911 & 0.920 \\ 
& + rand. VGG \cite{liu2021generic} & 32.823 & 0.872 & 2.344  & 0.552 & 34.233 &0.901 & 2.856 &0.831  & 0.913 & 0.921 \\ 
 \midrule
\parbox[t]{1mm}{\multirow{4}{*}{\rotatebox[origin=c]{90}{\scriptsize \emph{CoPer} loss}}} 
& + \emph{SimSiam}-ResNet (ours) & 32.910 & 0.879 & 2.310  & 0.568 \color{blue}{$\uparrow$ 5.1\%} & 34.301 &0.901 & 2.812 &0.830  & 0.912 & 0.920 \\ 
& + \emph{SimSiam}-VGG (ours) &  32.988 & 0.880 & 2.306  & 0.585 \color{blue}{$\uparrow$ 6.8\%} & 34.374 &0.902 & 2.786 &0.831  & 0.915 & 0.921 \\ 
& + \emph{SwAV}-ResNet (ours)& \textbf{32.997}  & {0.879} & \textbf{2.302} & 0.584 \color{red}{$\uparrow$ 6.7\%}  &  34.219 & 0.902 & 2.794 &0.831 & 0.915 &\textbf{0.922}    \\ 
& + \emph{SwAV}-VGG (ours)&  {32.980}  & \textbf{0.881}  &   {2.321}  & \textbf{0.613} \color{red}{$\uparrow$ 9.6\%} & \textbf{34.405}  & \textbf{0.903}  & \textbf{2.777} &\textbf{0.832} & \textbf{0.916} &\textbf{0.922}
 \\
 \midrule
& Upper bound & ~~-  & ~~-  &  ~~-  & 0.755 & -  & -  & - &0.863 & 0.936 &0.941
 \\
 \bottomrule
\end{tabular}
\vspace{-1mm}
\label{tab:quantitative_improvment}
\end{table*}

\label{sec:what_matter}
\subsection{What is relevant to capture perceptual similarity?}
We ask this question to understand why the contrastive methods behave differently in the \emph{2AFC} perceptual tests in Tab.~\ref{effectiveness}. We answer this question by correlating the accuracies in the \emph{2AFC} perceptual test to their top-1 accuracies on the \emph{ImageNet} which is an indicator of the representational power. 
We observe a strong correlation (Pearson's correlation coefficient, \textit{r = 0.761}, \textit{p = 0.028}) between them as shown in Fig.~ \ref{fig:scatter_plot}. This correlation allows us to objectively compare the abilities of the models to capture perceptual similarity. Considering this objective evaluation, we choose the \emph{SwAV} method \cite{caron2020unsupervised} to learn the domain-specific representation of medical data.

\subsection{Can the \emph{ImageNet}-derived representations be used as a perceptual metric for medical images?}
As explained before, we normalize the perceptual similarity computed by the perceptual metric and the ratings from two experts. In Fig.~\ref{fig:scatter_plot}, one can observe that the perceptual similarity measured by the \emph{CSR}-based metrics are closer to experts' ratings compared to the ones pre-trained on \emph{ImageNet}. This is expected because (1) an intrinsic domain shift is present between natural images and medical images, and (2) the self-supervised method is optimized on medical data. These findings are coherent with recent studies which discussed that self-supervised pre-training on medical data is more effective than directly borrowing the learned representations from \emph{ImageNet} \cite{zhou2019models,sowrirajan2021moco}. In the supplementary, we interpret the difference between the visual features captured by two kinds of models (i.e., supervised by \emph{ImageNet} and self-supervised by MR images) and visualize the feature maps of two \emph{VGGs} given an MR image as the input. 
Based on the quantitative analysis and qualitative observation, we conclude that the domain-specific \emph{CSR} can be a better perceptual metric compared to the representations learned from \emph{ImageNet}.

\subsection{Can the self-supervision-based perceptual loss enhance image quality?}
\begin{figure}[!htbp] 
\centering
\includegraphics[width=0.46\textwidth]{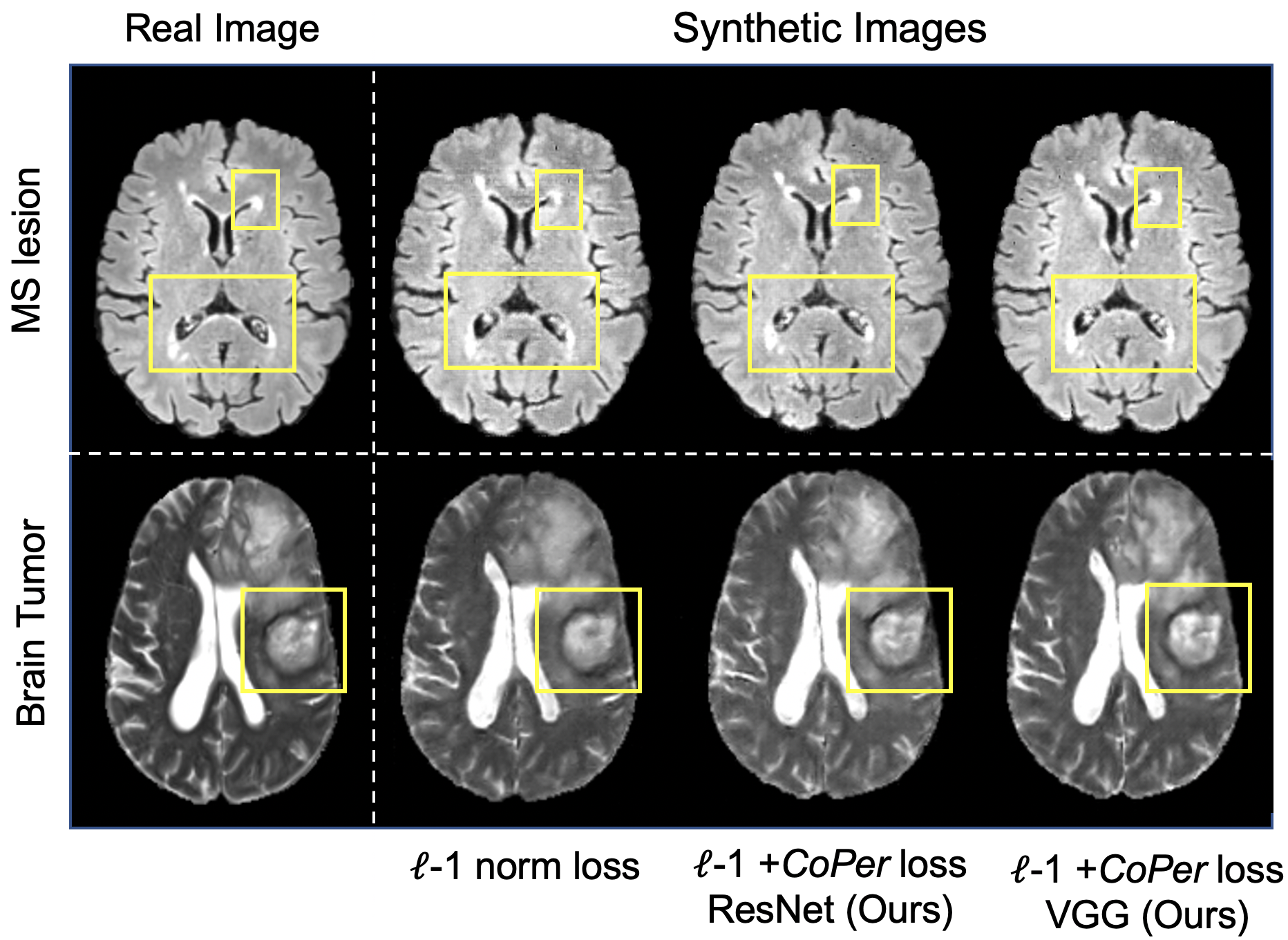}
\caption{\textbf{Sample synthetic images generated by different loss functions}: (1) an $\ell$1-norm loss, (2) a combination of $\ell$1-norm loss and our \emph{CoPer} loss with ResNet architecture, and (3) a combination of $\ell$1-norm loss and our \emph{CoPer} loss with VGG architecture. One can see that the images generated by solely using an $\ell$1-norm loss are relatively noisy and lack local textures of lesions and tumors.}
\label{fig:synthetic_images}
\end{figure}

To further explore the potential of the \emph{CSR}, we leverage it as a part of the loss function in image synthesis tasks, i.e., the \emph{CoPer} loss formulated in Methods section. We evaluate the image synthesis quality in two aspects: (1) direct quantitative metrics to assess the image quality and (2) indirect performance measure from the downstream automated segmentation task using the synthetic images. 
We use two traditional quality metrics (PSNR, SSIM) for direct quality comparison and the newly proposed \emph{CoPer} metric. Notably, these three metrics reflect pixel-wise error, structural similarity, and perceptual similarity, respectively. From Tab.~\ref{tab:quantitative_improvment}, we find that our \emph{CoPer} loss consistently enhances the pixel-wise {$\ell$-1 norm loss} in both image synthesis tasks across two architectures. Fig. \ref{fig:synthetic_images} shows the synthetic images generated by different loss functions. 
Interestingly, \emph{VGG} largely outperforms \emph{ResNet} in both image synthesis tasks, which is in line with existing work that \emph{VGG} is superior to \emph{ResNet} at capturing perceptual features in image regression tasks \cite{liu2021generic}. We further perform automated segmentation using pre-trained segmentation models \cite{isensee2021nnu,kofler2020brats} on the various sets of synthetic 3D scans generated by different loss functions. From Tab.~\ref{tab:quantitative_improvment}, we can see that the improved image quality boosts segmentation performance as a downstream task. Especially in the MS lesion dataset, it improves the Dice score over the baseline method by a large margin of 9.6\% when using a self-supervised VGG (Tab.~\ref{tab:quantitative_improvment}). Although randomized networks \cite{liu2021generic} can be used as a perceptual loss and improve the baseline {$\ell$-1 norm loss}, the improvement is marginal since they are not optimized on specific domains. We also compare \emph{SwAV} \cite{caron2020unsupervised} pre-training with another contrastive learning method named \emph{SimSiam} \cite{chen2020simple}. We observe that both methods consistently improve over the baselines while the \emph{SwAV} approaches learn better perceptual representation. 
\section{Discussion}
This work explores, for the first time, the perceptuality of contrastive self-supervised representation and benchmarks existing metrics and perceptual losses. In the natural image domain, we find that contrastive self-supervised representation captures comparable perceptuality to fully supervised ones. In the medical image domain, we show that domain-specific representations are better at capturing similarity than off-the-shelf representations pre-trained on \emph{ImageNet}. Our proposed new perceptual loss enhances image quality in two medical image synthesis tasks. Moreover, the improved image quality boosts downstream segmentation tasks. Our results suggest that the perceptuality from \textit{CSR} is an emergent property that could be adapted across many image domains, across architectures, and to 3D vision. 

One limitation of this work is that the number of training images is relatively small compared to \emph{ImageNet}. 
Although we demonstrate that contrastive self-supervision can be used as a domain-specific perceptual metric and improve the image quality as a loss function in image synthesis tasks, it is unclear how effective the learned representation is as there is no available evaluation metric (e.g., similar to top-1 accuracy on \emph{ImageNet}).
Due to the data heterogeneity (e.g., different acquisitions), training a self-supervised model on domains with diverse distributions \cite{feng2019self} is still an open problem. Another limitation is that the proposed \emph{CoPer} loss is performed in 2D and limits its use to 2D medical images. Considering the availability of 3D neuroimaging data, the extension to 3D would be straightforward and naturally requires contrastive self-supervised pre-training in a 3D manner \cite{sanghi2020info3d,chaitanya2020contrastive,taleb20203d,zeng2021positional,dufumier2021contrastive}. We will investigate this direction in future work.

\section{Appendix}
\subsection{Computing the perceptual similarity with ViT.}
\begin{figure}[t!]
\centering
\includegraphics[width=0.48\textwidth]{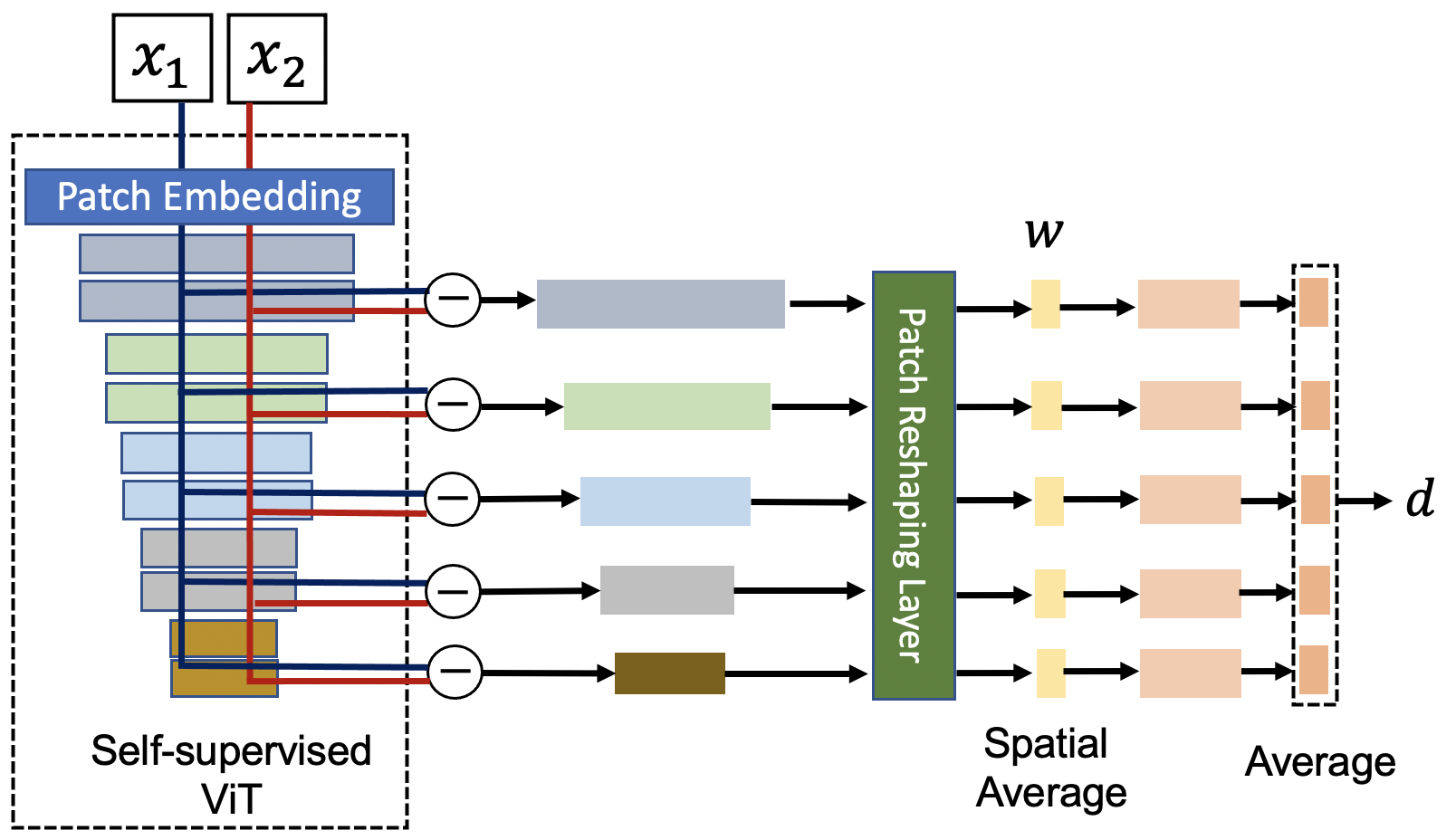}
\vspace{-1mm}
\caption{\textbf{Computing the perceptual similarity given two images and a Vision Transformer (ViT)}. Different from the \emph{CNN} architecture, an input image is cut into patches as an embedding. To compute the distance \emph{d} between two images, x$_{1}$, x$_{2}$, given a pre-trained \emph{ViT}, we first compute deep feature maps, compute the difference, reshape the two-dimension tensor (M$\times$HW) to three-dimensional (M$\times$H$\times$W), and scale each channel. We then average across spatial dimensions and all layers.}
\label{fig:dino_distance}
\end{figure}

The main text presents how to compute the perceptual similarity given two images and a \emph{CNN} architecture. Here we present the extension of the method to Vision Transformer (ViT), corresponding to the result of the \emph{DINO} method in Table 2 of the main manuscript. 
Figure \ref{fig:dino_distance} presents the modified version of the process to calculate perceptual similarity given two images and one pre-trained ViT (i.e., \emph{DINO} \cite{caron2021emerging}). An input image is passed through a 16 $\times$ 16 Conv layer with 16 $\times$ 16 strides which was referred as \emph{Patch Embedding} layer. The five activations through five blocks of \emph{DINO}. Each \emph{DINO} block consists of a normalization layer, a self-attention layer, and a fully-connected layer proposed in the original model \cite{caron2021emerging}. Each block has a fixed number of output channels fixed at 384. To be consistent with the idea of spatial-pooling of features, we pass the activations to a \emph{Patch Reshaping Layer} that reshapes the output tensor to a shape of 384 $\times$ 14 $\times$ 14. Generally, given a pre-trained \emph{ViT}, to compute the distance \emph{d} between two images, x$_{1}$, x$_{2}$, we first compute deep feature maps, compute the difference, reshape the two-dimension tensor (M$\times$HW) to three-dimensional (M$\times$H$\times$W), and scale each channel. We then average across spatial dimensions and all layers. We used the ViT-small model (ViT-S/16) in the experiments.

\subsection{Training setting of \emph{SwAV}.} \label{cont_train}

We used the \emph{SwAV} method \cite{caron2020unsupervised} for self-supervised pre-training on the 2D slices of two MRI datasets: MS lesion and brain tumor. For both tasks, we train the models for 400 epochs with a batch size of 96 using one GPU with 24 GB memory. We use \emph{SGD} and cosine annealing for learning rate (LR) scheduling. The base LR is set to 0.6 and final LR is 0.0006. The warm-up epoch is fixed at 5.

\begin{figure}
\centering
\includegraphics[width=0.45\textwidth]{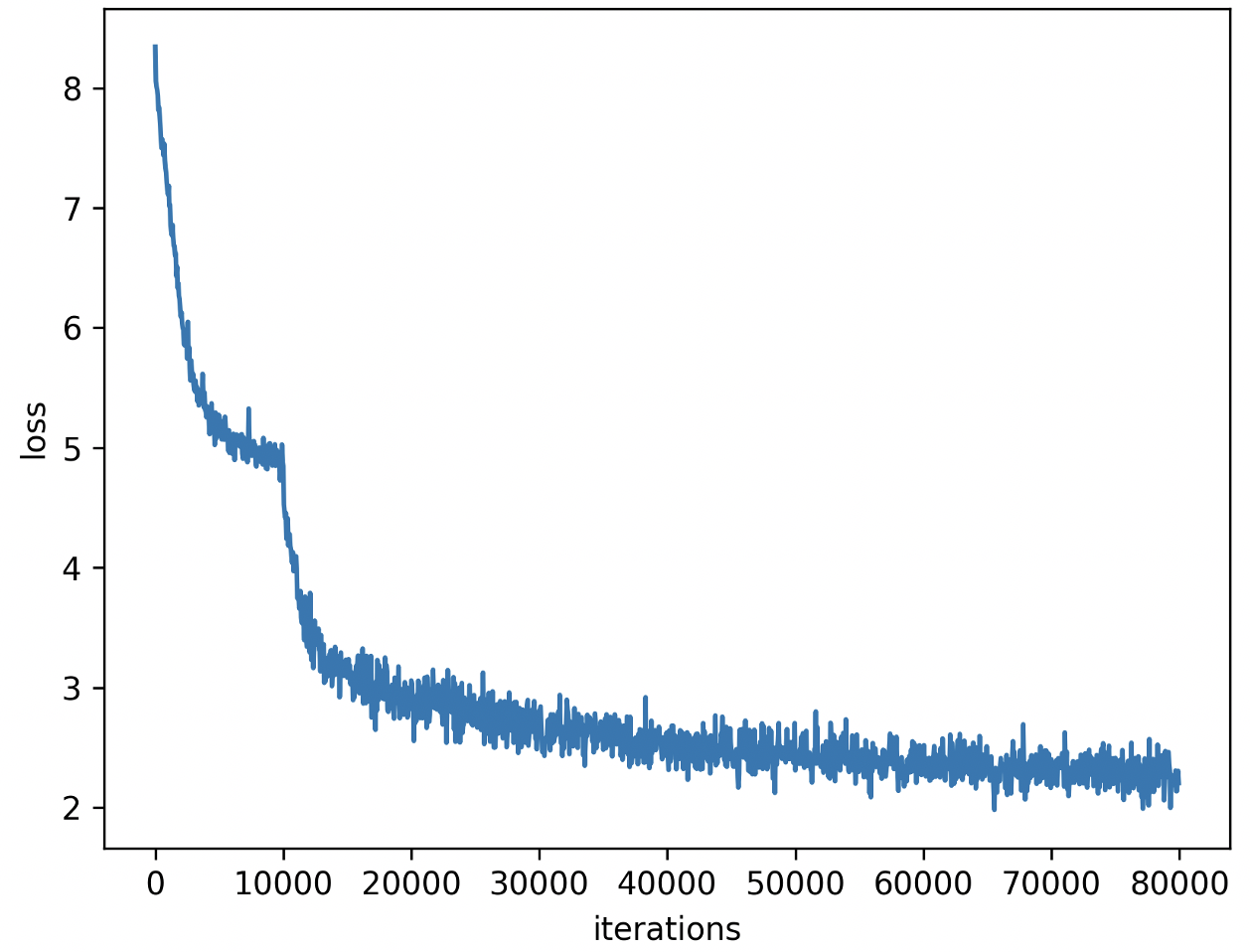}
\vspace{-1mm}
\caption{\textbf{Training loss of over the iterations (400 epochs in total).} Each epoch contains around 187 iterations for \emph{18k} samples with a batch size of 96. The loss values were logged every 50 iterations.}
\label{fig:training}
\end{figure}

For each gray-scale 2D slice image, we stack it with its two copies to be a three-channel image. We then applied two random crops of sizes 192 and 96. For the crop size of 192, the scale is varied between a minimum of 0.14 to a maximum of 1.0; for the image crop of size 96, the scale is varied between a minimum of 0.05 and a maximum of 0.14. For the clustering, 331 iterations are performed to obtain the prototypes across the training. Along with random cropping, additional transformations such as random horizontal flipping (with a probability of 0.5), color jitter (with a probability of 0.8), brightness adjustment with a factor of 0.8, contrast adjustment with a factor of 0.8 and saturation with a factor of 0.8), and random Gaussian blurring. 
We use \emph{51k} samples for pre-training on the MS lesion dataset and \emph{88k} samples for the brain tumor dataset.
Figure \ref{fig:training} shows the training process of 400 epochs (122100 iterations) on the MS lesion dataset.


\begin{figure*}[t!]
\centering
\includegraphics[width=0.98\textwidth]{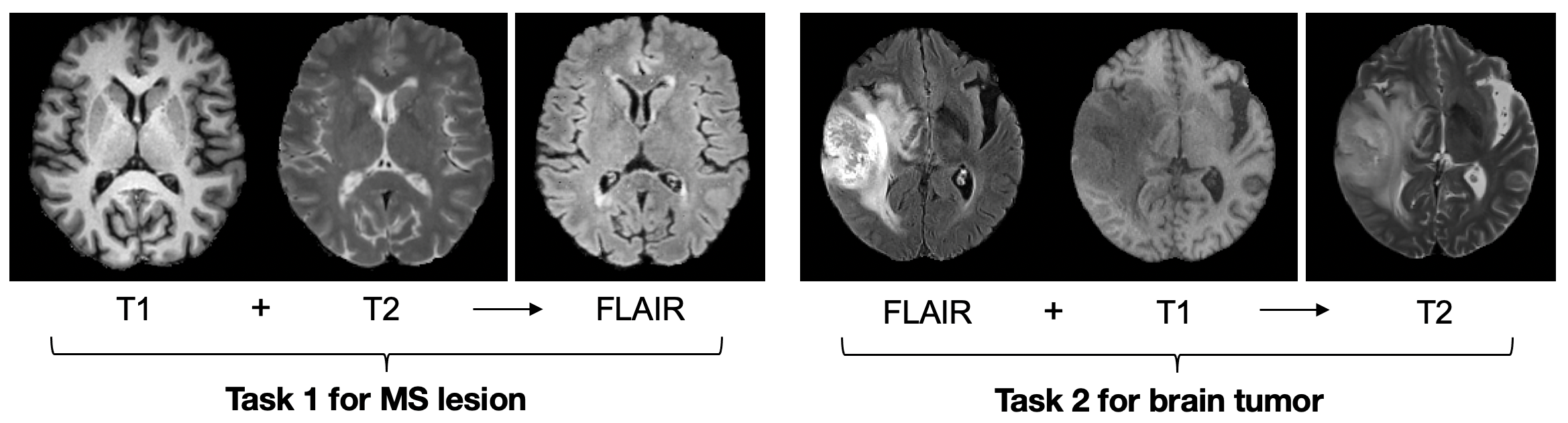}
\caption{\textbf{Two image synthesis tasks for MS lesion and brain tumor MRI scans.}. Task~1 aims to reconstruct the FLAIR sequence given the combination of T1 and T2. FLAIR and T1 are standard sequences to detect and segment MS lesions in the brains \cite{valverde2017improving}. This task is to study if a FLAIR image can be synthesized by using T1 and T2 when it is of bad quality. Task 2 aims to synthesize a missing T2 sequence to facilitate tumor treatment. In clinical practice, the T2 sequence is sometimes not imaged due to the treatment urgency.}
\label{fig:tasks}
\end{figure*}

\subsection{Training setting of two image synthesis tasks.}
We extend the image-to-image translation method \emph{pix2pix}  \cite{isola2017image} to a multi-modal version for our image synthesis tasks. The learning rate is set at 0.0002. The images are resized to 286 x 286, and random crops of size 256 x 256 are taken during training while we use a fixed image size of 256 $\times$ 256 at inference. The images are flipped horizontally with a probability of 0.5 and normalized with a mean of 0.5 and a standard deviation of 0.5. 
For the MS lesion dataset, 4351 training samples were used for training, while for the brain tumor images, 11969 training samples were used. The best-performing models were selected by observing the training loss and the evaluation metric using an internal validation set which includes 10\% of the training set. 

\begin{figure}[t!]
\centering
\includegraphics[width=0.48\textwidth]{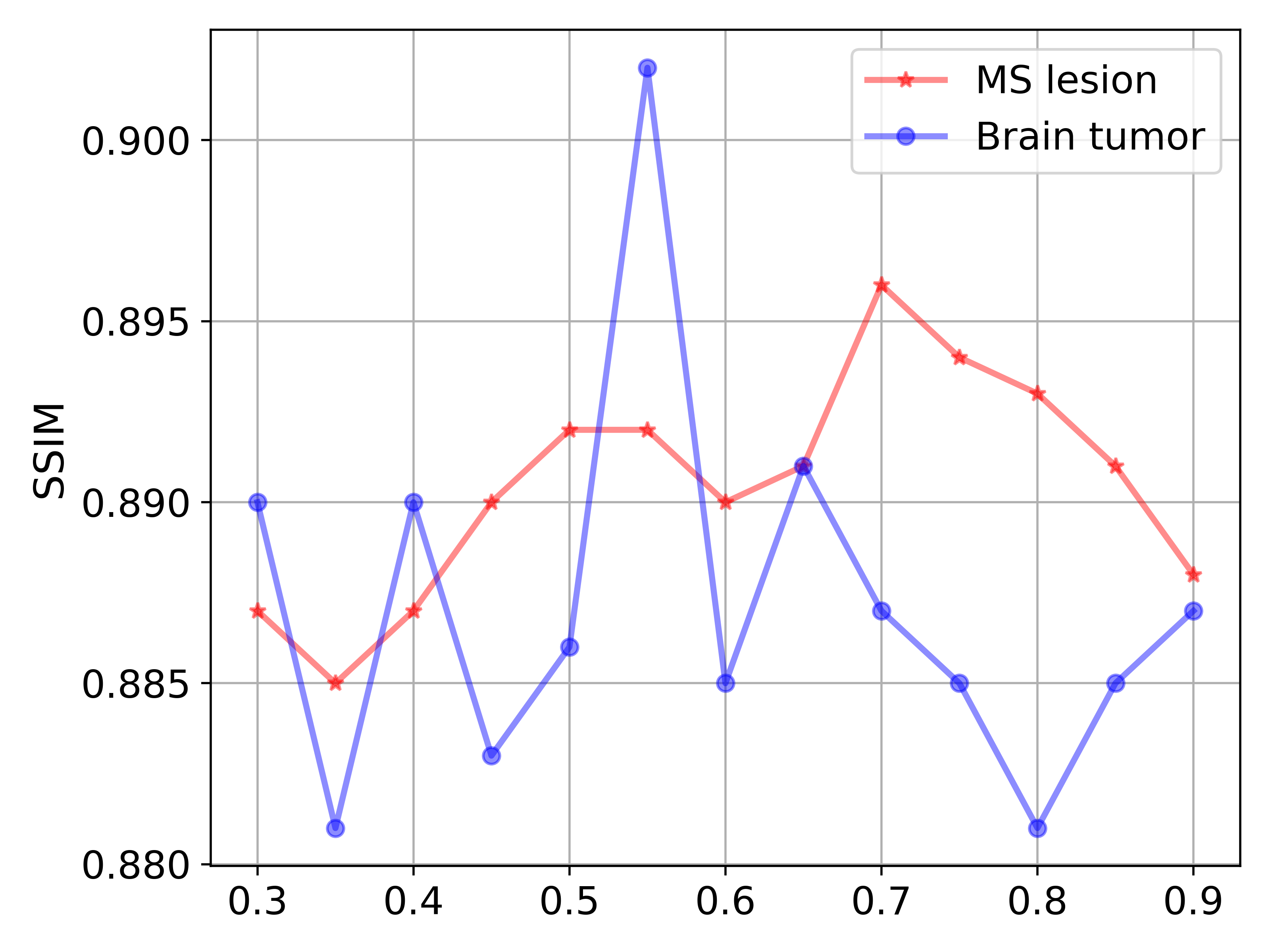}
\vspace{-1mm}
\caption{The effect of $\lambda_{1}$ to balance the pixel-wise loss and perceptual loss in the image translation task. We observe that $\lambda_{1}$ can be specifically optimized for the different translation tasks.}
\label{fig:lambda}
\end{figure}

\subsection{Domain shift between the representations learned from different image domains.}
We investigate the domain specificity of the representations learned from different imaging modalities, i.e., representation shift between different imaging modalities. To illustrate the domain shift between them, we transfer the representations from one task to another and quantitatively compare the image quality of synthetic images in two `cross-perceptuality' synthesis tasks. Specifically, for the MS lesion FLAIR image synthesis task, we transfer the features learned from brain tumor T2 images, serving as a perceptual loss and vice versa. As we can observe from Table~\ref{tab:perceptual_transfer}, the perceptual transfers from other medical sub-domains degrade the image quality of the translated target domains in two synthesis tasks. In contrast, it is interesting that the pre-trained \emph{VGG} (on \emph{ImageNet}) does not harm but improves the image quality in the medical image translation tasks. This leaves us questioning what a good perceptual representation would be given a specific medical image domain. Although the representation shift exists in heterogeneous medical data (even just among different MR modalities), the potential of contrastive self-supervision is demonstrated in this work.  

\begin{table}[htbp]
\centering
\setlength{\tabcolsep}{4pt}
\begin{tabular}{lccccccc}
\specialrule{.1em}{0em}{.1em}
\multicolumn{1}{c}{\multirow{2}{*}{\textbf{Methods}}} & \multicolumn{3}{c}{MS lesion}  & \multicolumn{3}{c}{ Brain tumor} \\ \cline{2-7}  & {PSNR} & {SSIM} & CoPer $\downarrow$ & \multicolumn{1}{l}{PSNR} & {SSIM} & CoPer $\downarrow$\\ \specialrule{.1em}{0em}{.1em}
$\ell$1+supervised VGG & 32.992 & 0.885  &  2.343  & 34.307 & 0.901 & 2.780
\\
$\ell$1+\emph{CoPer loss}-VGG &  {32.980}  & {0.881}  &   {2.321}  & {34.405}  & {0.903}  & {2.777}

\\
\specialrule{.1em}{0em}{.1em}
$\ell$1 + transferred loss 1 &  \cellcolor{red!10}{32.770}  & \cellcolor{red!10}{0.877} & \cellcolor{red!10}{2.460}  & - & - & - 

\\
$\ell$1 + transferred loss 2 &  -  &  - & - & \cellcolor{red!10}{34.231} & \cellcolor{red!10}{0.899} & \cellcolor{red!10}{2.829}
                                            
 \\
\specialrule{.1em}{0em}{.1em}
\end{tabular}
\caption{\textbf{Intra-domain perceptual transfers degrade image quality.} \emph{transferred loss 1} denotes using the perceptual features from brain tumor \emph{T2} images and applying it for MS lesion \emph{FLAIR} synthesis task; \emph{transferred loss 2} denotes using the perceptual features from MS lesion \emph{FLAIR} images and applying it for brain tumor \emph{T2} synthesis task. We observed that the perceptual transfer from different medical sub-domains degrade the image quality of the 
translated target domains.}
\label{tab:perceptual_transfer}
\end{table}


\subsection{The effect of hyper-parameters.}
We test the effect of $\lambda_{1}$, which balances pixel-wise and feature-based loss. In Fig. \ref{fig:lambda}, we observe that $\lambda_{1}$ should be optimized in a separate validation set. The best $\lambda_{1}$ for the MS lesion and the brain tumor translation tasks were 0.55 and 0.70, respectively. The best value could be obtained on an independent validation set.  We used different $\lambda_{1}$ for different architectures in the comparison experiments.

\begin{figure*}[t!]
\centering
\includegraphics[width=0.9\textwidth]{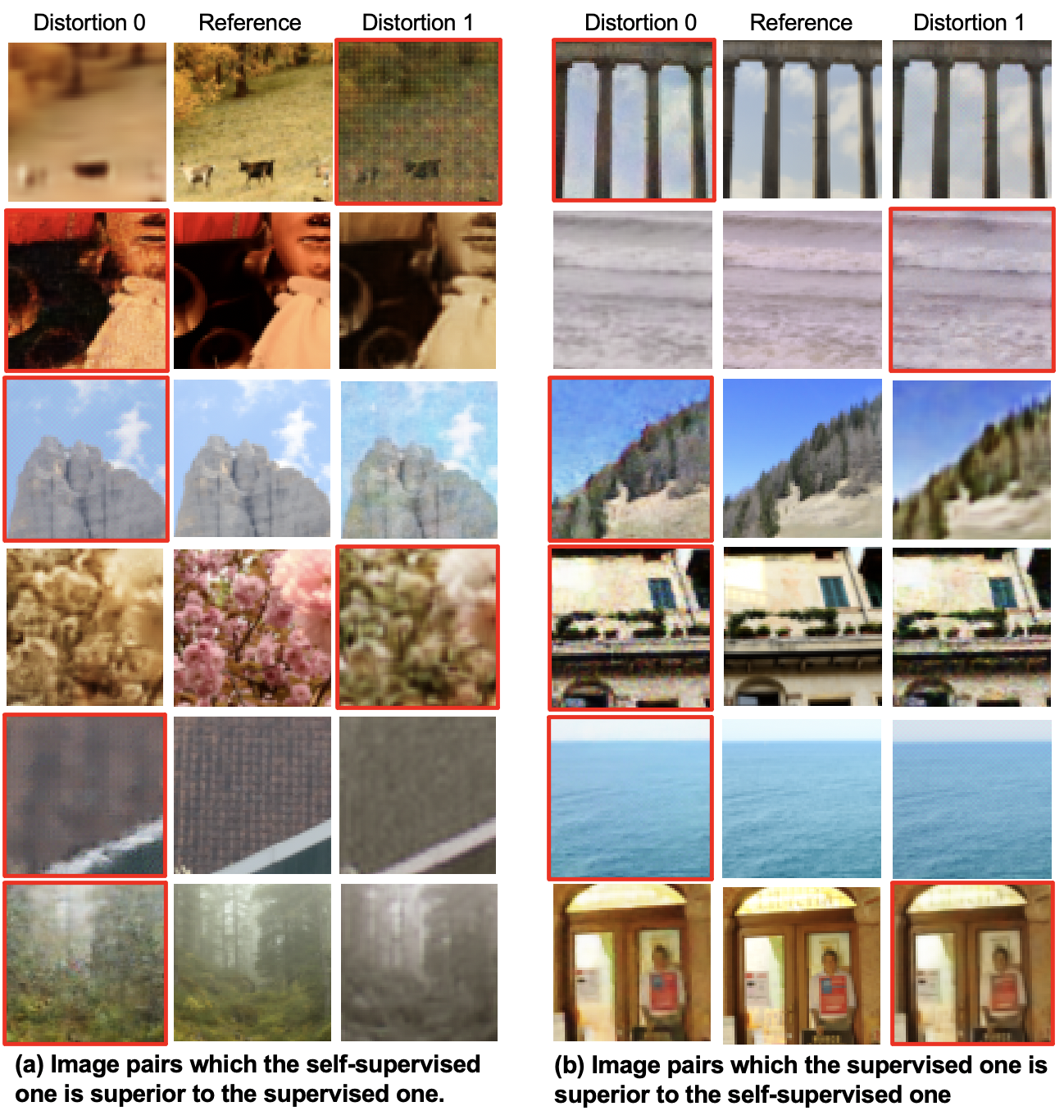}
\caption{\textbf{Samples of images where one method (self-supervised or supervised) has the same judgments as humans while the other does not.}  We observed that the \emph{DINO} one seems to be robust to global perceptual distortion (rows 1, 4, 5, and 6 in Fig.~\ref{fig:samples_comparison}a) while the supervised \emph{ResNet-50} is sensitive to small local distortions (rows 1 and 6 in Fig.~\ref{fig:samples_comparison}b).}
\label{fig:samples_comparison}
\end{figure*}


\subsection{On what images does the self-supervised one outperform supervised ones in perceptual ratings?}
To investigate the behavior of the self-supervised method and supervised method, we display the image pairs in which the two methods (i.e., \emph{DINO} \cite{caron2021emerging} and supervised \emph{ResNet-50} \cite{he2016deep}) make different judgments (based on the relative perceptual similarity). Fig.~\ref{fig:samples_comparison} shows the sample results. We observed that the \emph{DINO} method seems to be robust to global perceptual distortion (rows 1, 4, 5, 6 in Fig.~\ref{fig:samples_comparison}a) while the supervised \emph{ResNet-50} is sensitive to small local distortions (rows 1 and 6 in Fig.~\ref{fig:samples_comparison}b). 

\subsection{Additional synthetic images.}

\begin{figure*}[t!]
\centering
\includegraphics[width=0.98\textwidth]{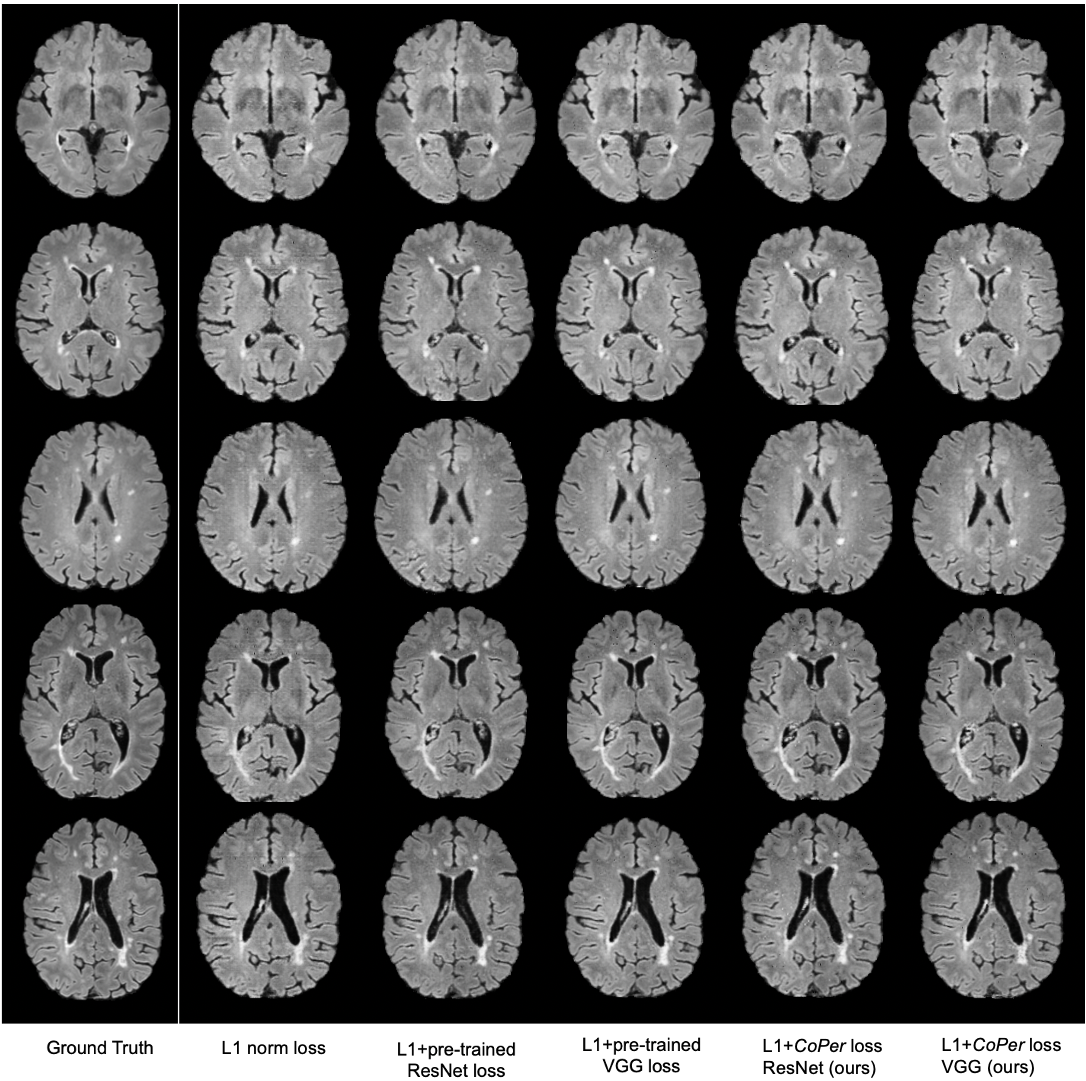}
\caption{\textbf{Synthetic FLAIR images of MS lesion patients generated by different loss functions}: 1) an $\ell$1-norm loss, 2) a combination of $\ell$1-norm loss and the loss using pre-trained \emph{ResNet} and 3) a combination of $\ell$1-norm loss and the loss using pre-trained \emph{VGG}, 4) a combination of $\ell$1-norm loss and our \emph{CoPer} loss with ResNet architecture and 5) a combination of $\ell$1-norm loss and our \emph{CoPer} loss with VGG architecture. One can see that the images generated by solely using an $\ell$1-norm loss are relatively noisy and lack details of local structures, such as lesions, while the perceptual loss can significantly enhance the image quality.}
\label{fig:ms_images}
\end{figure*}

\begin{figure*}[t!]
\centering
\includegraphics[width=0.98\textwidth]{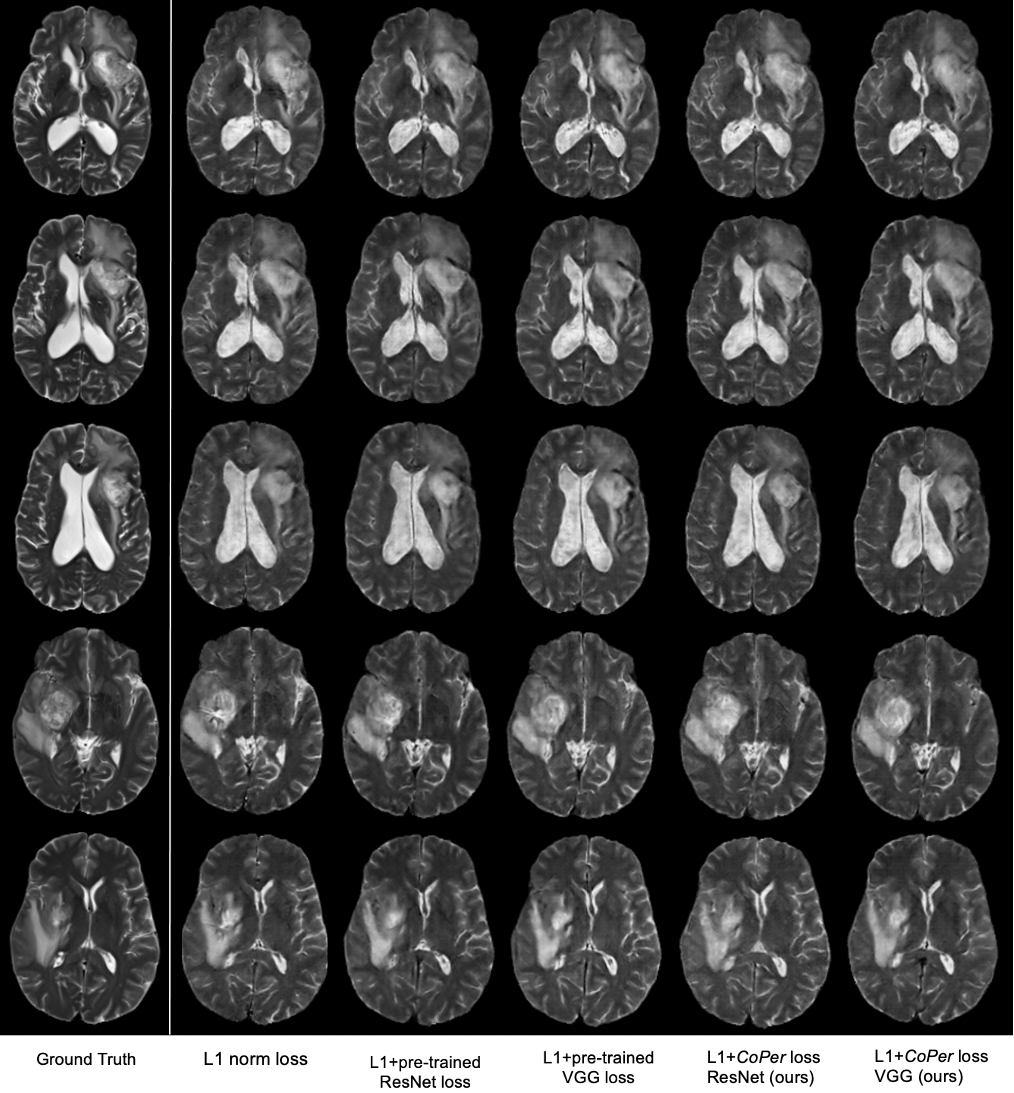}
\caption{\textbf{Synthetic T2 images of brain tumor patients generated by different loss functions}: 1) an $\ell$1-norm loss, 2) a combination of $\ell$1-norm loss and the loss using pre-trained \emph{ResNet} and 3) a combination of $\ell$1-norm loss and the loss using pre-trained \emph{VGG}, 4) a combination of $\ell$1-norm loss and our \emph{CoPer} loss with ResNet architecture and 5) a combination of $\ell$1-norm loss and our \emph{CoPer} loss with VGG architecture. One can see that the images generated by solely using an $\ell$1-norm loss are relatively noisy with unrealistic texture, while the perceptual loss can significantly enhance the image quality of brain tumor region. }
\label{fig:tumor_images}
\end{figure*}

Figure \ref{fig:ms_images} and Figure \ref{fig:tumor_images} show the additional synthetic images generated by five loss functions (including the pre-trained \emph{VGG} and \emph{ResNet-50}) of MS lesion and brain tumor patients, respectively. We can observe that the proposed perceptual loss enhances the quality of relevant local structures, such as lesions or tumors.

\clearpage

\clearpage

\bibliographystyle{IEEEtran}
\bibliography{reference.bib}

\begin{thebibliography}{10}
\providecommand{\url}[1]{#1}
\csname url@samestyle\endcsname
\providecommand{\newblock}{\relax}
\providecommand{\bibinfo}[2]{#2}
\providecommand{\BIBentrySTDinterwordspacing}{\spaceskip=0pt\relax}
\providecommand{\BIBentryALTinterwordstretchfactor}{4}
\providecommand{\BIBentryALTinterwordspacing}{\spaceskip=\fontdimen2\font plus
\BIBentryALTinterwordstretchfactor\fontdimen3\font minus
  \fontdimen4\font\relax}
\providecommand{\BIBforeignlanguage}[2]{{%
\expandafter\ifx\csname l@#1\endcsname\relax
\typeout{** WARNING: IEEEtran.bst: No hyphenation pattern has been}%
\typeout{** loaded for the language `#1'. Using the pattern for}%
\typeout{** the default language instead.}%
\else
\language=\csname l@#1\endcsname
\fi
#2}}
\providecommand{\BIBdecl}{\relax}
\BIBdecl

\bibitem{zhang2018unreasonable}
R.~Zhang, P.~Isola, A.~A. Efros, E.~Shechtman, and O.~Wang, ``The unreasonable
  effectiveness of deep features as a perceptual metric,'' in \emph{Proceedings
  of the IEEE conference on computer vision and pattern recognition}, 2018, pp.
  586--595.

\bibitem{wang2004image}
Z.~Wang, A.~C. Bovik, H.~R. Sheikh, and E.~P. Simoncelli, ``Image quality
  assessment: from error visibility to structural similarity,'' \emph{IEEE
  transactions on image processing}, vol.~13, no.~4, pp. 600--612, 2004.

\bibitem{simonyan2014very}
K.~Simonyan and A.~Zisserman, ``Very deep convolutional networks for
  large-scale image recognition,'' \emph{arXiv preprint arXiv:1409.1556}, 2014.

\bibitem{gatys2016image}
L.~A. Gatys, A.~S. Ecker, and M.~Bethge, ``Image style transfer using
  convolutional neural networks,'' in \emph{Proceedings of the IEEE conference
  on computer vision and pattern recognition}, 2016, pp. 2414--2423.

\bibitem{johnson2016perceptual}
J.~Johnson, A.~Alahi, and L.~Fei-Fei, ``Perceptual losses for real-time style
  transfer and super-resolution,'' in \emph{European conference on computer
  vision}.\hskip 1em plus 0.5em minus 0.4em\relax Springer, 2016, pp. 694--711.

\bibitem{amir2021understanding}
D.~Amir and Y.~Weiss, ``Understanding and simplifying perceptual distances,''
  in \emph{Proceedings of the IEEE/CVF Conference on Computer Vision and
  Pattern Recognition}, 2021, pp. 12\,226--12\,235.

\bibitem{dosovitskiy2016generating}
A.~Dosovitskiy and T.~Brox, ``Generating images with perceptual similarity
  metrics based on deep networks,'' \emph{Advances in neural information
  processing systems}, vol.~29, pp. 658--666, 2016.

\bibitem{pan2021hyperspectral}
E.~Pan, Y.~Ma, F.~Fan, X.~Mei, and J.~Huang, ``Hyperspectral image
  classification across different datasets: A generalization to unseen
  categories,'' \emph{Remote Sensing}, vol.~13, no.~9, p. 1672, 2021.

\bibitem{deng2009imagenet}
J.~Deng, W.~Dong, R.~Socher, L.-J. Li, K.~Li, and L.~Fei-Fei, ``Imagenet: A
  large-scale hierarchical image database,'' in \emph{2009 IEEE conference on
  computer vision and pattern recognition}.\hskip 1em plus 0.5em minus
  0.4em\relax Ieee, 2009, pp. 248--255.

\bibitem{he2016deep}
K.~He, X.~Zhang, S.~Ren, and J.~Sun, ``Deep residual learning for image
  recognition,'' in \emph{Proceedings of the IEEE conference on computer vision
  and pattern recognition}, 2016, pp. 770--778.

\bibitem{liu2021generic}
Y.~Liu, H.~Chen, Y.~Chen, W.~Yin, and C.~Shen, ``Generic perceptual loss for
  modeling structured output dependencies,'' in \emph{Proceedings of the
  IEEE/CVF Conference on Computer Vision and Pattern Recognition}, 2021, pp.
  5424--5432.

\bibitem{wang2020cliffnet}
L.~Wang, J.~Zhang, Y.~Wang, H.~Lu, and X.~Ruan, ``Cliffnet for monocular depth
  estimation with hierarchical embedding loss,'' in \emph{European Conference
  on Computer Vision}.\hskip 1em plus 0.5em minus 0.4em\relax Springer, 2020,
  pp. 316--331.

\bibitem{caron2020unsupervised}
M.~Caron, I.~Misra, J.~Mairal, P.~Goyal, P.~Bojanowski, and A.~Joulin,
  ``Unsupervised learning of visual features by contrasting cluster
  assignments,'' \emph{arXiv preprint arXiv:2006.09882}, 2020.

\bibitem{he2020momentum}
K.~He, H.~Fan, Y.~Wu, S.~Xie, and R.~Girshick, ``Momentum contrast for
  unsupervised visual representation learning,'' in \emph{Proceedings of the
  IEEE/CVF Conference on Computer Vision and Pattern Recognition}, 2020, pp.
  9729--9738.

\bibitem{chen2020simple}
T.~Chen, S.~Kornblith, M.~Norouzi, and G.~Hinton, ``A simple framework for
  contrastive learning of visual representations,'' in \emph{International
  conference on machine learning}.\hskip 1em plus 0.5em minus 0.4em\relax PMLR,
  2020, pp. 1597--1607.

\bibitem{caron2021emerging}
M.~Caron, H.~Touvron, I.~Misra, H.~J{\'e}gou, J.~Mairal, P.~Bojanowski, and
  A.~Joulin, ``Emerging properties in self-supervised vision transformers,''
  \emph{arXiv preprint arXiv:2104.14294}, 2021.

\bibitem{zbontar2021barlow}
J.~Zbontar, L.~Jing, I.~Misra, Y.~LeCun, and S.~Deny, ``Barlow twins:
  Self-supervised learning via redundancy reduction,'' \emph{arXiv preprint
  arXiv:2103.03230}, 2021.

\bibitem{azizi2021big}
S.~Azizi, B.~Mustafa, F.~Ryan, Z.~Beaver, J.~Freyberg, J.~Deaton, A.~Loh,
  A.~Karthikesalingam, S.~Kornblith, T.~Chen \emph{et~al.}, ``Big
  self-supervised models advance medical image classification,'' \emph{arXiv
  preprint arXiv:2101.05224}, 2021.

\bibitem{chaitanya2020contrastive}
K.~Chaitanya, E.~Erdil, N.~Karani, and E.~Konukoglu, ``Contrastive learning of
  global and local features for medical image segmentation with limited
  annotations,'' \emph{arXiv preprint arXiv:2006.10511}, 2020.

\bibitem{taleb20203d}
A.~Taleb, W.~Loetzsch, N.~Danz, J.~Severin, T.~Gaertner, B.~Bergner, and
  C.~Lippert, ``3d self-supervised methods for medical imaging,'' \emph{arXiv
  preprint arXiv:2006.03829}, 2020.

\bibitem{zeng2021positional}
D.~Zeng, Y.~Wu, X.~Hu, X.~Xu, H.~Yuan, M.~Huang, J.~Zhuang, J.~Hu, and Y.~Shi,
  ``Positional contrastive learning for volumetricmedical image segmentation,''
  \emph{arXiv preprint arXiv:2106.09157}, 2021.

\bibitem{madhusudana2022image}
P.~C. Madhusudana, N.~Birkbeck, Y.~Wang, B.~Adsumilli, and A.~C. Bovik, ``Image
  quality assessment using contrastive learning,'' \emph{IEEE Transactions on
  Image Processing}, vol.~31, pp. 4149--4161, 2022.

\bibitem{li2019diamondgan}
H.~Li, J.~C. Paetzold, A.~Sekuboyina, F.~Kofler, J.~Zhang, J.~S. Kirschke,
  B.~Wiestler, and B.~Menze, ``Diamondgan: unified multi-modal generative
  adversarial networks for mri sequences synthesis,'' in \emph{International
  Conference on Medical Image Computing and Computer-Assisted
  Intervention}.\hskip 1em plus 0.5em minus 0.4em\relax Springer, 2019, pp.
  795--803.

\bibitem{kofler2021we}
F.~Kofler, I.~Ezhov, F.~Isensee, F.~Balsiger, C.~Berger, M.~Koerner,
  J.~Paetzold, H.~Li, S.~Shit, R.~McKinley \emph{et~al.}, ``Are we using
  appropriate segmentation metrics? identifying correlates of human expert
  perception for cnn training beyond rolling the dice coefficient,''
  \emph{arXiv preprint arXiv:2103.06205}, 2021.

\bibitem{isola2017image}
P.~Isola, J.-Y. Zhu, T.~Zhou, and A.~A. Efros, ``Image-to-image translation
  with conditional adversarial networks,'' in \emph{Proceedings of the IEEE
  conference on computer vision and pattern recognition}, 2017, pp. 1125--1134.

\bibitem{zhu2017unpaired}
J.-Y. Zhu, T.~Park, P.~Isola, and A.~A. Efros, ``Unpaired image-to-image
  translation using cycle-consistent adversarial networks,'' in
  \emph{Proceedings of the IEEE international conference on computer vision},
  2017, pp. 2223--2232.

\bibitem{qasim2020red}
A.~B. Qasim, I.~Ezhov, S.~Shit, O.~Schoppe, J.~C. Paetzold, A.~Sekuboyina,
  F.~Kofler, J.~Lipkova, H.~Li, and B.~Menze, ``Red-gan: Attacking class
  imbalance via conditioned generation. yet another medical imaging
  perspective.'' in \emph{Medical Imaging with Deep Learning}.\hskip 1em plus
  0.5em minus 0.4em\relax PMLR, 2020, pp. 655--668.

\bibitem{yurt2021mustgan}
M.~Yurt, S.~U. Dar, A.~Erdem, E.~Erdem, K.~K. Oguz, and T.~{\c{C}}ukur,
  ``Mustgan: Multi-stream generative adversarial networks for mr image
  synthesis,'' \emph{Medical Image Analysis}, vol.~70, p. 101944, 2021.

\bibitem{thomas2021improving}
M.~F. Thomas, F.~Kofler, L.~Grundl, T.~Finck, H.~Li, C.~Zimmer, B.~Menze, and
  B.~Wiestler, ``Improving automated glioma segmentation in routine clinical
  use through artificial intelligence-based replacement of missing sequences
  with synthetic magnetic resonance imaging scans.'' \emph{Investigative
  Radiology}, 2021.

\bibitem{mason2019comparison}
A.~Mason, J.~Rioux, S.~E. Clarke, A.~Costa, M.~Schmidt, V.~Keough, T.~Huynh,
  and S.~Beyea, ``Comparison of objective image quality metrics to expert
  radiologists’ scoring of diagnostic quality of mr images,'' \emph{IEEE
  transactions on medical imaging}, vol.~39, no.~4, pp. 1064--1072, 2019.

\bibitem{bakas2018identifying}
S.~Bakas, M.~Reyes, A.~Jakab, S.~Bauer, M.~Rempfler, A.~Crimi, R.~T. Shinohara,
  C.~Berger, S.~M. Ha, M.~Rozycki \emph{et~al.}, ``Identifying the best machine
  learning algorithms for brain tumor segmentation, progression assessment, and
  overall survival prediction in the brats challenge,'' \emph{arXiv preprint
  arXiv:1811.02629}, 2018.

\bibitem{menze2014multimodal}
B.~H. Menze, A.~Jakab, S.~Bauer, J.~Kalpathy-Cramer, K.~Farahani, J.~Kirby,
  Y.~Burren, N.~Porz, J.~Slotboom, R.~Wiest \emph{et~al.}, ``The multimodal
  brain tumor image segmentation benchmark (brats),'' \emph{IEEE transactions
  on medical imaging}, vol.~34, no.~10, pp. 1993--2024, 2014.

\bibitem{chen2021exploring}
X.~Chen and K.~He, ``Exploring simple siamese representation learning,'' in
  \emph{Proceedings of the IEEE/CVF Conference on Computer Vision and Pattern
  Recognition}, 2021, pp. 15\,750--15\,758.

\bibitem{hore2010image}
A.~Hore and D.~Ziou, ``Image quality metrics: Psnr vs. ssim,'' in \emph{2010
  20th international conference on pattern recognition}.\hskip 1em plus 0.5em
  minus 0.4em\relax IEEE, 2010, pp. 2366--2369.

\bibitem{zhang2016colorful}
R.~Zhang, P.~Isola, and A.~A. Efros, ``Colorful image colorization,'' in
  \emph{European conference on computer vision}.\hskip 1em plus 0.5em minus
  0.4em\relax Springer, 2016, pp. 649--666.

\bibitem{noroozi2016unsupervised}
M.~Noroozi and P.~Favaro, ``Unsupervised learning of visual representations by
  solving jigsaw puzzles,'' in \emph{European conference on computer
  vision}.\hskip 1em plus 0.5em minus 0.4em\relax Springer, 2016, pp. 69--84.

\bibitem{zhou2019models}
Z.~Zhou, V.~Sodha, M.~M.~R. Siddiquee, R.~Feng, N.~Tajbakhsh, M.~B. Gotway, and
  J.~Liang, ``Models genesis: Generic autodidactic models for 3d medical image
  analysis,'' in \emph{International conference on medical image computing and
  computer-assisted intervention}.\hskip 1em plus 0.5em minus 0.4em\relax
  Springer, 2019, pp. 384--393.

\bibitem{sowrirajan2021moco}
H.~Sowrirajan, J.~Yang, A.~Y. Ng, and P.~Rajpurkar, ``Moco pretraining improves
  representation and transferability of chest x-ray models,'' in \emph{Medical
  Imaging with Deep Learning}.\hskip 1em plus 0.5em minus 0.4em\relax PMLR,
  2021, pp. 728--744.

\bibitem{isensee2021nnu}
F.~Isensee, P.~F. Jaeger, S.~A. Kohl, J.~Petersen, and K.~H. Maier-Hein,
  ``nnu-net: a self-configuring method for deep learning-based biomedical image
  segmentation,'' \emph{Nature methods}, vol.~18, no.~2, pp. 203--211, 2021.

\bibitem{kofler2020brats}
F.~Kofler, C.~Berger, D.~Waldmannstetter, J.~Lipkova, I.~Ezhov, G.~Tetteh,
  J.~Kirschke, C.~Zimmer, B.~Wiestler, and B.~H. Menze, ``Brats toolkit:
  translating brats brain tumor segmentation algorithms into clinical and
  scientific practice,'' \emph{Frontiers in neuroscience}, vol.~14, p. 125,
  2020.

\bibitem{feng2019self}
Z.~Feng, C.~Xu, and D.~Tao, ``Self-supervised representation learning from
  multi-domain data,'' in \emph{Proceedings of the IEEE/CVF International
  Conference on Computer Vision}, 2019, pp. 3245--3255.

\bibitem{sanghi2020info3d}
A.~Sanghi, ``Info3d: Representation learning on 3d objects using mutual
  information maximization and contrastive learning,'' in \emph{European
  Conference on Computer Vision}.\hskip 1em plus 0.5em minus 0.4em\relax
  Springer, 2020, pp. 626--642.

\bibitem{dufumier2021contrastive}
B.~Dufumier, P.~Gori, J.~Victor, A.~Grigis, M.~Wessa, P.~Brambilla, P.~Favre,
  M.~Polosan, C.~McDonald, C.~M. Piguet \emph{et~al.}, ``Contrastive learning
  with continuous proxy meta-data for 3d mri classification,'' \emph{arXiv
  preprint arXiv:2106.08808}, 2021.

\end{thebibliography}

\end{document}